\documentclass[fleqn,usenatbib]{mnras}
\usepackage{newtxtext,newtxmath}

\usepackage[T1]{fontenc}

\usepackage{amsmath}
\usepackage{multirow}
\usepackage[dvips]{graphicx}
\usepackage{hyperref}
\usepackage{cleveref}
\usepackage{threeparttable}
\usepackage[dvipsnames]{xcolor}
\usepackage{ulem}

\graphicspath{{./}{figures/}}

\title[MOJAVE XX. Persistent polarization in AGN jets]{MOJAVE XX. Persistent Linear Polarization Structure in Parsec-scale AGN Jets}

\author[Pushkarev et al.]{A.~B.~Pushkarev,$^{1,2}$\thanks{E-mail: pushkarev.alexander@gmail.com}
H.~D.~Aller,$^{3}$
M.~F.~Aller,$^{3}$
D.~C.~Homan,$^{4}$
Y.~Y.~Kovalev,$^{5,2,6}$
M.~L.~Lister,$^{7}$
\newauthor
I.~N.~Pashchenko,$^{2}$
T.~Savolainen,$^{8,9,5}$
D.~I.~Zobnina$^{2,6}$\\
$^{1}$Crimean Astrophysical Observatory, 298409 Nauchny, Crimea, Russia\\
$^{2}$Astro Space Centre of Lebedev Physical Institute, Profsoyuznaya 84/32, Moscow 117997, Russia\\
$^{3}$Department of Astronomy, University of Michigan, Ann Arbor, MI 48109-1107 USA\\
$^{4}$Department of Physics and Astronomy, Denison University, Granville, OH 43023, USA\\
$^{5}$Max-Planck-Institut f\"ur Radioastronomie, Auf dem H\"ugel 69, D-53121 Bonn, Germany\\
$^{6}$Moscow Institute of Physics and Technology, Institutsky per. 9, Dolgoprudny, Moscow region, 141700, Russia\\
$^{7}$Department of Physics and Astronomy, Purdue University, 525 Northwestern Avenue, West Lafayette, IN 47907, USA\\
$^{8}$Aalto University Department of Electronics and Nanoengineering, PL 15500, FI-00076 Aalto, Finland\\
$^{9}$Aalto University Mets\"ahovi Radio Observatory, Mets\"ahovintie 114, FI-02540 Kylm\"al\"a, Finland
}

\date{
Accepted 2023 February 09. Received 2023 February 09; in original form 2022 September 12 
}
\pubyear{2023}

\begin{document}
\label{firstpage}
\pagerange{\pageref{firstpage}--\pageref{lastpage}}
\maketitle

\begin{abstract}
We analysed the parsec-scale linear polarization properties of 436 active galactic nuclei (AGN) based on 15~GHz polarimetric Very Long Baseline Array (VLBA) observations. We present polarization and total intensity images averaged over at least five epochs since 1996 January 19 through 2019 August 4. Stacking improves the image sensitivity down to $\sim$30~$\mu$Jy beam$^{-1}$ and effectively fills out the jet cross-section both in total intensity and linear polarization. It delineates the long-term persistent magnetic field configuration and its regularity by restoring spatial distributions of the electric vector position angle (EVPA) and fractional polarization, respectively. On average, about ten years of stacking period is needed to reveal the stable and most-complete polarization distribution of a source. We find that the degree of polarization significantly increases down and across the jet towards its edges, typically manifesting U or W-shaped transverse profiles, suggesting a presence of a large-scale helical magnetic field associated with the outflow. In some AGN jets, mainly BL Lacs, we detect quasi-constant fractional polarization profiles across the jet, accompanied by EVPAs that closely follow the outflow. BL Lacs show higher fractional polarization values in their cores and jets than those in quasars up to hectoparsec de-projected scales, while on larger scales, they become comparable. High-synchrotron-peaked BL Lac jets are found to be less polarized than intermediate and low-synchrotron-peaked BL Lacs. The spatial distribution of the EVPAs in BL Lacs tend to align with the local jet direction, while quasars show an excess of orthogonal polarization orientation.
\end{abstract}


\section{Introduction} 
\label{sec:intro}
Active galactic nuclei (AGN) are the most powerful non-transient sources in the Universe. Their centers host supermassive black holes (SMBH). It is widely accepted that bi-polar antiparallel outflows are launched as a result of the accretion of matter on an SMBH, with the extraction of energy and angular momentum from the rotating central machine \citep{BZ77} and/or accretion disk \citep{BP82}. Simulations show that the most efficient outflows are produced if spins of the SMBH and accretion disk are prograde \citep{Tchekhovskoy12}. AGN jets emit radiation over the entire electromagnetic spectrum, from radio to GeV--TeV gamma rays. Recently, evidence was found that radio-bright blazars are the sources of high-energy astrophysical neutrinos \citep[e.g.][see also discussion in \citealt{Zhou21} and \citealt{Plavin22}]{Plavin20,Plavin21,Hovatta2021,Buson22}.

The jets propagate through the ambient medium, which is a factor of $10^2$--$10^3$ times more dense than the jet \citep{Marti19}. Changes in the ISM pressure profile form a series of quasi-stationary recollimation shocks in the jet predicted by theory, at least for super-magnetosonic jets \citep{Meier11,Mizuno15} and also detected with very long baseline interferometry (VLBI) observations as slow pattern speed components in the inner jet regions, mainly within 3--4~pc projected distance from the radio core \citep{Cohen14,Jorstad17,MOJAVE_XVII}. The outflows initially have a quasi-parabolic shape \citep{Vlahakis03,Beskin06,Komissarov07} and undergo a transition to quasi-conical geometry at distances of $10^5$--$10^6$ gravitational radii, most likely due to a change from magnetically dominated to an equipartition regime, as found for a dozen of nearby ($z<0.07$) AGN for which VLBI provides high enough linear resolution \citep{Asada12,Pushkarev_2017,Kovalev20}. Post-recollimation-shock(s) jet features are superluminal, with apparent speeds up to $\sim$40$c$ \citep{Jorstad05,Piner12,MOJAVE_XVIII} showing signatures of residual acceleration up to the de-projected distances of $\sim$100~pc, beyond which the outflows either manifest a constant speed regime or start to decelerate \citep{MOJAVE_XII}.

The key agent in forming, accelerating and collimating the relativistic jets is the magnetic field originally threaded through the accretion disc and then propelled out and wound up in the outflow. An indication for a toroidal component to the magnetic field in a number of transversely resolved AGN jets was found, as evident from the gradients of Faraday rotation measure across the outflow \citep[e.g.][]{Asada02,ZT05,Hovatta_2012,Zamaninasab13,Gabuzda15,2017MNRAS.467...83K}. Toroidal magnetic fields imply that a current flows in the jet, with the expected values in a range of $10^{17}$--$10^{18}$~A \citep{Wardle18}. On the other hand, high variability in the flux and linear polarization of AGN jets suggests the presence of shocks and plasma turbulence, so that the magnetic field is partly disordered being a combination of regular and chaotic magnetic field components \citep{Marscher_2014,Marscher_2017}. In this paper, we study the stable component of the magnetic field of AGN jets, its regularity and configuration by constructing and analysing fractional polarization and electric vector position angle (EVPA) distributions averaged over the available epochs of VLBA observations. In the companion paper, \cite{MOJAVE_XXI} based on the same observational data, we investigated the decade-long variability of the linear polarization of parsec-scale AGN outflows.

The layout of the paper is as follows. In \autoref{sec:obs}, we describe our observational data and the properties of the studied AGN jet sample. We discuss our method of constructing stacked maps in total intensity and linear polarization in \autoref{sec:stacked_maps} and present the statistics of the images. We present our results in \autoref{sec:p_prop} and summarize the main conclusions in \autoref{sec:summary}. By the term `core' we mean the apparent origin of AGN jets, which is typically seen as the brightest and most compact feature in VLBI images of blazars \citep[e.g.][]{Lobanov98}. The spectral index $\alpha$ is defined as $S_\nu\propto\nu^\alpha$, where $S_\nu$ is the flux density measured at frequency $\nu$. All position angles are given in degrees east of north. We adopt a cosmology with $\Omega_m=0.27$, $\Omega_\Lambda=0.73$ and $H_0=71$~km~s$^{-1}$~Mpc$^{-1}$ \citep{Komatsu09}.

\section{Observational data and source sample} 
\label{sec:obs}
For the purposes of our study, we selected sources that have at least five polarization-sensitive observing epochs with the VLBA at 15~GHz. Over 90~per~cent of the observations were performed within the MOJAVE program\footnote{\url{https://www.cv.nrao.edu/MOJAVE}}, while the rest were taken from the NRAO archive. We excluded the quasar 2023+335, which is subject to refractive-dominated scattering~\citep{Pushkarev13} resulting in the unclear position of the core component in a number of observing epochs. The quasar 1329$-$126 initially having five epochs was also dropped as one epoch had poor data quality. Thus, our sample comprises 436 sources listed in \autoref{t:sources}, 207 of which are members of the MOJAVE 1.5~Jy quarter-century (QC) flux-density limited sample \citep{MOJAVE_XVIII} containing 232 objects. Four epochs of 0316+413 prior to 1999 December 27 were excluded due to the unclear core position of the source. Overall, we have 5921 single-epoch maps at 368 unique observational epochs between 1996 January 19 to 2019 August 4.

\begin{figure}
    \centering
    \includegraphics[width=\linewidth]{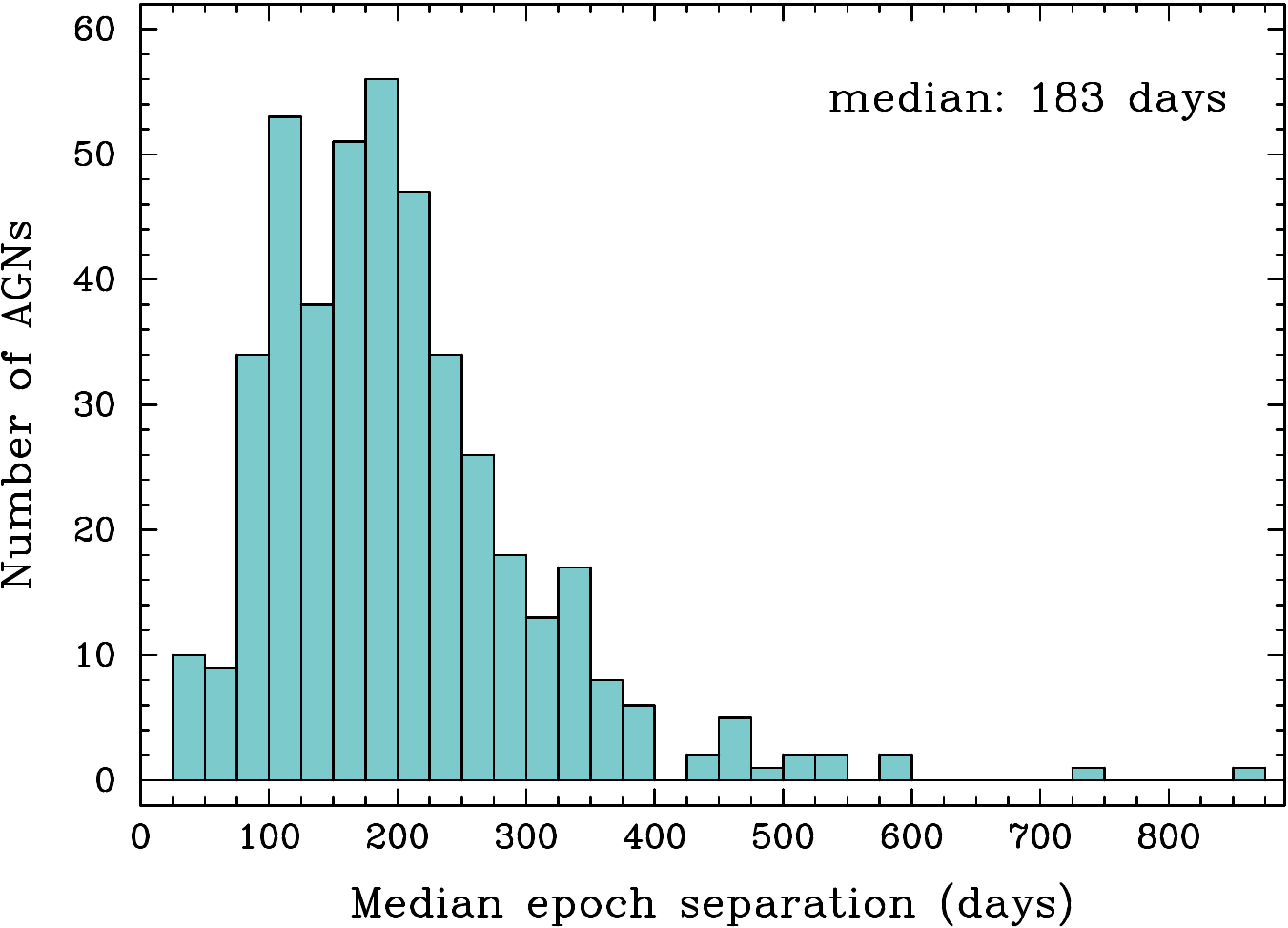}
    \caption{Histogram of typical time sampling for 436 sources having at least five polarimetric observing epochs. 
    \label{f:cadence}
    }
\end{figure}

The sources in the MOJAVE program are observed with individual cadence determined by the rate of the corresponding morphological changes. The distribution of the median time separation of the neighbouring epochs for the 436 sources from our sample is wide (\autoref{f:cadence}), ranging from 38 days for the most frequently observed source, BL Lacertae object 0716+714, to 861 days for the radio galaxy 1345+125. The median cadence for the whole sample is six months.

Additional details on the source sample, including time coverage, redshift, optical and SED class distributions, are provided in the companion paper by \cite{MOJAVE_XXI}. The data reduction, the process of single-epoch imaging and  structure model fitting are described in \citet{Lister_2009,Lister_2018}.

\begin{table*}
\centering
\caption{Source properties.
\label{t:sources}}
\begin{threeparttable}
\begin{tabular}{ccllccccr}
\hline\hline\noalign{\smallskip}
B1950.0 name & J2000.0 name & Alias & {$z$} & Opt. class & MOJAVE 1.5 & LAT & Spectrum & References \\
(1) & (2) & (3) & (4) & (5) & (6) & (7) & (8)  & (9) \\
\hline
0003$-$066 & J0006$-$0623 &       NRAO 005 & 0.3467 & B & Y & Y & LSP & \cite{2005PASA...22..277J},2 \\
0003$+$380 & J0005$+$3820 &     S4 0003+38 & 0.229  & Q & N & Y & LSP & \cite{1994AAS..103..349S},1 \\
0006$+$061 & J0009$+$0628 &   TXS 0006+061 & \ldots & B & N & Y & LSP & \cite{2012AA...538A..26R},1 \\
0007$+$106 & J0010$+$1058 &       III Zw 2 & 0.0893 & G & Y & Y & LSP & \cite{1970ApJ...160..405S},3 \\
0010$+$405 & J0013$+$4051 &      4C +40.01 & 0.256  & Q & N & Y & LSP & \cite{1992ApJS...81....1T},2 \\
0011$+$189 & J0013$+$1910 &  RGB J0013+191 & 0.477  & B & N & Y & LSP & \cite{2013ApJ...764..135S},2 \\
0012$+$610 & J0014$+$6117 &      4C +60.01 & \ldots & U & N & Y & LSP & \ldots,1 \\
0014$+$813 & J0017$+$8135 &    S5 0014+813 & 3.382  & Q & N & N & LSP & \cite{1987AZh....64..262V},3 \\
0015$-$054 & J0017$-$0512 & PMN J0017-0512 & 0.226  & Q & N & Y & LSP & \cite{2012ApJ...748...49S},1 \\
0016$+$731 & J0019$+$7327 &     S5 0016+73 & 1.781  & Q & Y & Y & LSP & \cite{1986AJ.....91..494L},2 \\
\hline
\end{tabular}
\begin{tablenotes}
\item
Columns are as follows:
(1) Source name in B1950.0 coordinates;
(2) Source name in J2000.0 coordinates;
(3) Alias;
(4) Redshift;
(5) Optical Class (Q=quasar, B=BL~Lac, G=radio galaxy, N=narrow-line Seyfert 1, U=unknown);
(6) Member of the MOJAVE 1.5~Jy QC Sample (Y = yes, N = no);
(7) Fermi LAT-detected;
(8) SED Class (LSP/ISP/HSP = Low/Intermediate/High Synchrotron Peaked);
(9) References for Redshift/Optical and SED Classification.
SED property references are as follows:
 1 = \cite{2015ApJ...810...14A},
 2 = \cite{4LAC},
 3 = \cite{2011arXiv1103.0749S},
 4 = \cite{2011ApJ...740...98M},
 5 = \cite{2015MNRAS.450.3568X},
 6 = \cite{2017AA...598A..17C},
 7 = \cite{2008AA...488..867N},
 8 = \cite{2017ApJS..232...18A},
 9 = \cite{2011ApJ...743..171A},
10 = \cite{2009ApJ...707L.142A},
11 = \cite{2006AA...445..441N},
12 = \cite{3HSP},
13 = \cite{2009ApJ...707...55A}, and
14 = \cite{2015AA...578A..69H}.
(This table is available in its entirety in a machine-readable form as supplementary material and at the CDS VizieR.)
\end{tablenotes}
\end{threeparttable}
\end{table*}

\section{Total intensity and linear polarization stacked images} 
\label{sec:stacked_maps}

To produce a stacked map in total intensity, we performed the following steps. First, every single-epoch $I$ image was shifted such that the core component, whose position is determined from the structure model fitting and assumed to be stable (though small non-zero shifts caused by the emergence of new components are possible), was placed at the phase centre of the map. Second, the source brightness distribution represented by the CLEAN components was convolved with a circular restoring beam, the full width at half maximum (FWHM) of which in milliarcseconds is the average between
\begin{equation}
1.283 - 8.950\cdot10^{-3}\delta - 7.914\cdot10^{-5}\delta^2 + 1.245\cdot10^{-6}\delta^3
\label{e:bmaj_fit}
\end{equation}
and
\begin{equation}
0.522 + 1.007\cdot10^{-3}\delta + 8.884\cdot10^{-6}\delta^2 - 5.571\cdot10^{-8}\delta^3\,,
\label{e:bmin_fit}
\end{equation}
where $\delta$ is the source J2000 declination in degrees, \autoref{e:bmaj_fit} and \autoref{e:bmin_fit} are the spline fits to the maximum and minimum FWHM dimensions of the naturally weighted elliptical restoring beam for several thousand 15~GHz VLBA observations made within the MOJAVE program. Thereafter, all single-epochs $I$ maps were averaged to make a stacked image $\overline{I}$.

As for the polarization stacking, there are two possible methods. First, used in this paper, is to follow the same procedure as for Stokes I but for Stokes Q and U, and then produce stacked maps of linearly polarized intensity $\overline{P}=\sqrt{\overline{Q}^2+\overline{U}^2}$, fractional polarization $m=\overline{P}/\overline{I}$ and its direction $\mathrm{EVPA}=0.5\,\mathrm{atan}(\overline{U}/\overline{Q})$. This approach treats polarization as a vector. An alternative method, applied in \cite{MOJAVE_XXI} is based on making single-epoch polarization maps first and then taking their scalar mean (regardless of EVPA) to derive typical values, e.g. to study variability.

Following \cite{Wardle74}, we corrected the stacked $P$ maps for the Ricean bias, i.e. $P_\mathrm{true}\approx P_\mathrm{obs}\sqrt{1-(\sigma_P/P_\mathrm{obs})^2}$. The corrections are $\lesssim3.2$ per cent for the regions above the lowest polarization contour drawn at the $4\sigma_P$ level, where $\sigma_P^2$ was estimated as 
\begin{equation}
\sum_{i,j=1}^N P_{ij}^2/(2N)\,,
\end{equation}
where $0<P_{ij}<5\sigma_I$, as the noise statistics of the $P$ map in a sky region, which either blank or having a low $P$ signal-to-noise ratio, follow a Rayleigh distribution. In turn, $\sigma_I$ was calculated using the following procedure: (i) for each outer quadrant of the map (with a square of 1/16 of a total image), the rms level is calculated, (ii) the highest rms value is removed and (iii) the median over three remaining rms estimates is taken.
The estimations of uncertainty in $I$, $P$, EVPA and polarization feed leakage derived from the simulations and analytical expressions are discussed in Appendix~\ref{a:errors}.
       
Finally, we corrected the total and linear polarization intensity distributions for the CLEANing bias (Appendix~\ref{a:clean_bias}). After subtracting the $P$-bias, we show only positive values of polarization intensity and corresponding fractional polarization. This is justified as (i) fewer than 1~per~cent of all pixels with $P>4\sigma_P$ presented in the stacked images are negative, i.e.\ $P_{\rm stack}-P_{\rm bias}<0$ and (ii) fewer than 1~per~cent of the negative $P$-pixels are significantly (at $>3\sigma$ level) different from zero.

We also filtered out epochs with poor data quality. If the noise level of a single-epoch total intensity map exceeded three times the corresponding median rms for a given source, the epoch was excluded. In total, we ruled out 65 noisy maps (about 1~per~cent of the total number) from 31 sources. This additionally increased the dynamic range of the stacked images by a factor up to a few, especially in the cases with a small number of the observing epochs.

\begin{figure*}
    \centering
    \includegraphics[width=0.49\linewidth]{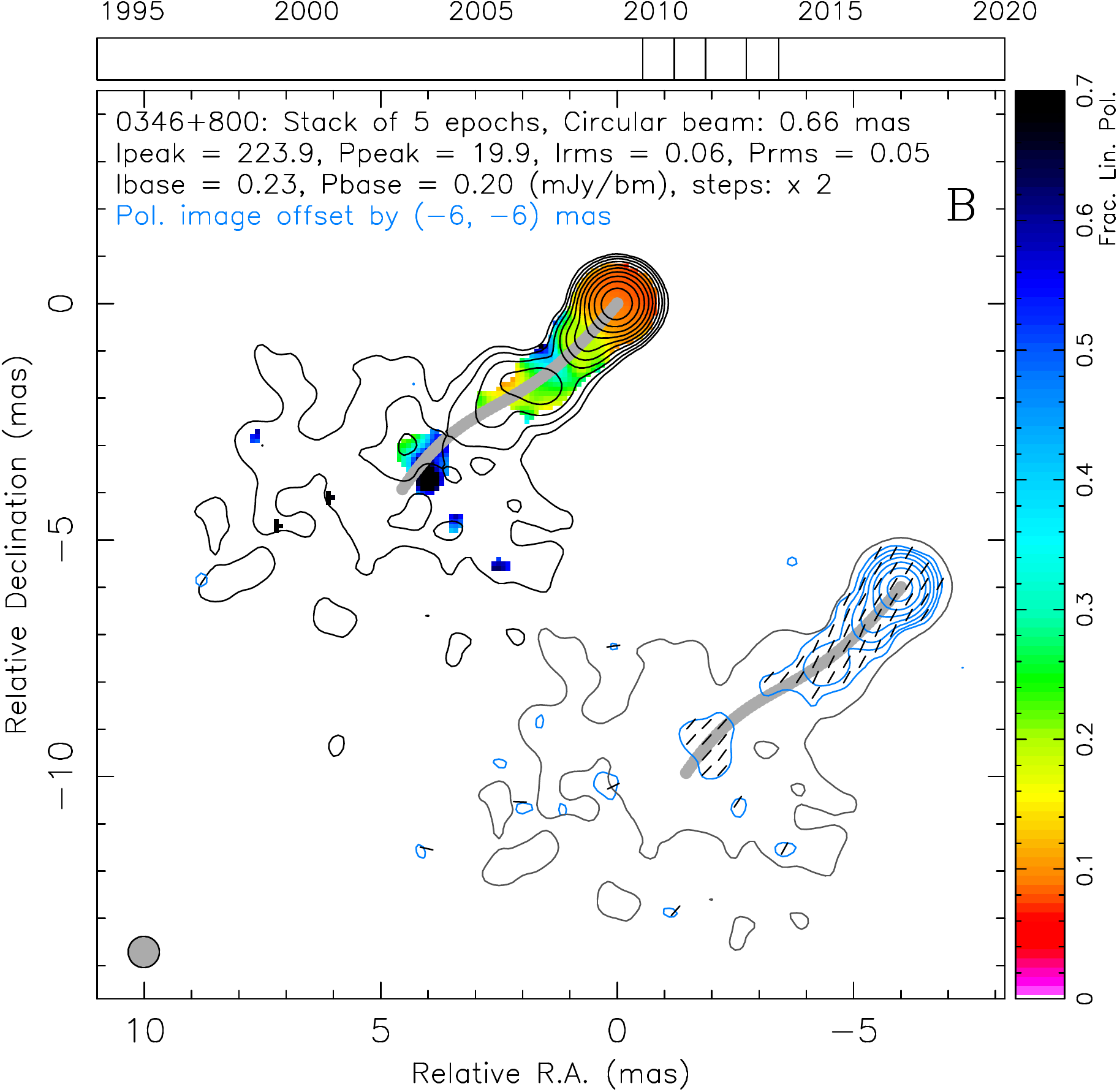}\hspace{0.1cm}
    \includegraphics[width=0.49\linewidth]{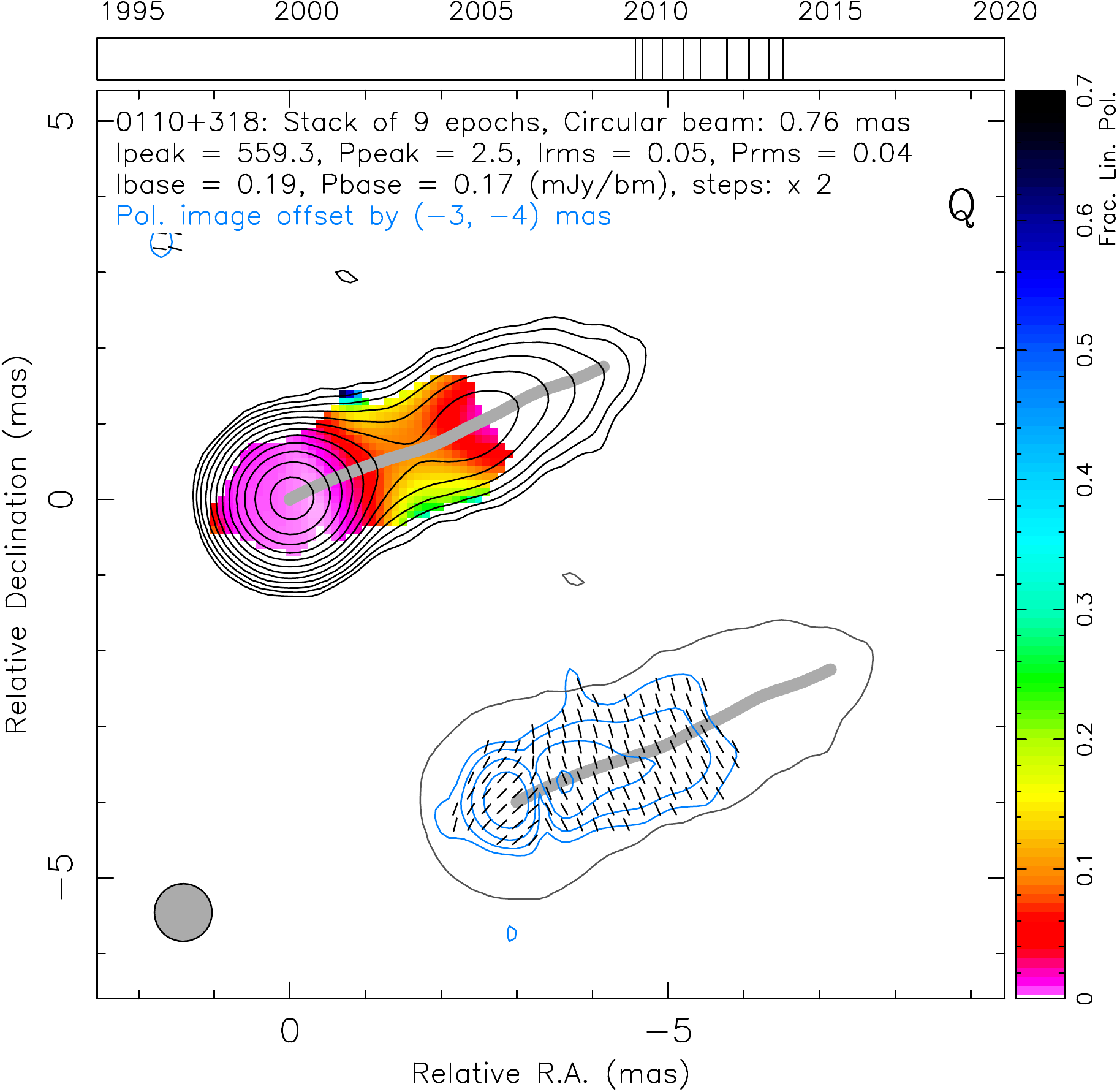}\vspace{0.5cm}
    \includegraphics[width=0.49\linewidth]{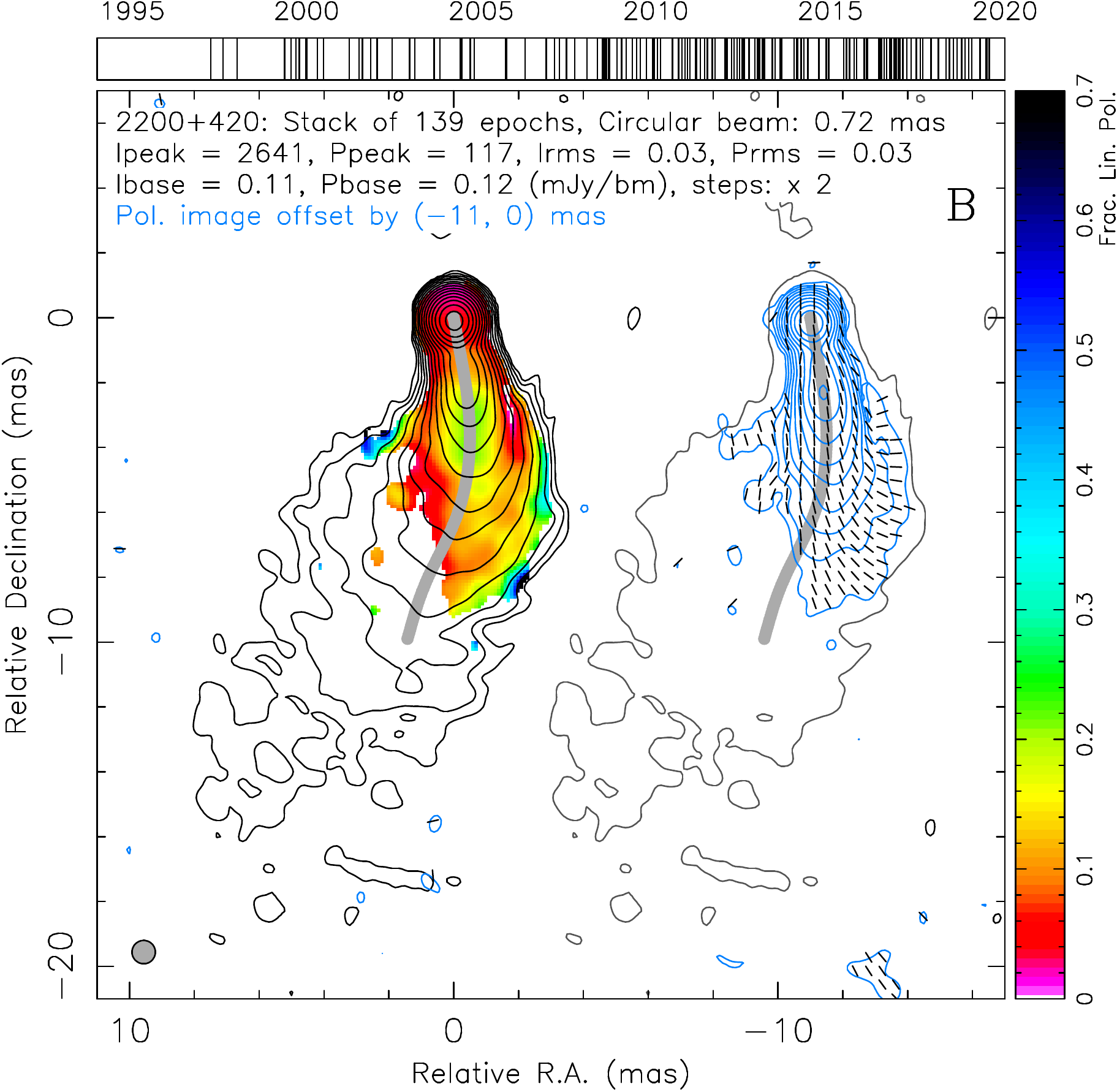}\vspace{0.5cm}
    \includegraphics[width=0.49\linewidth]{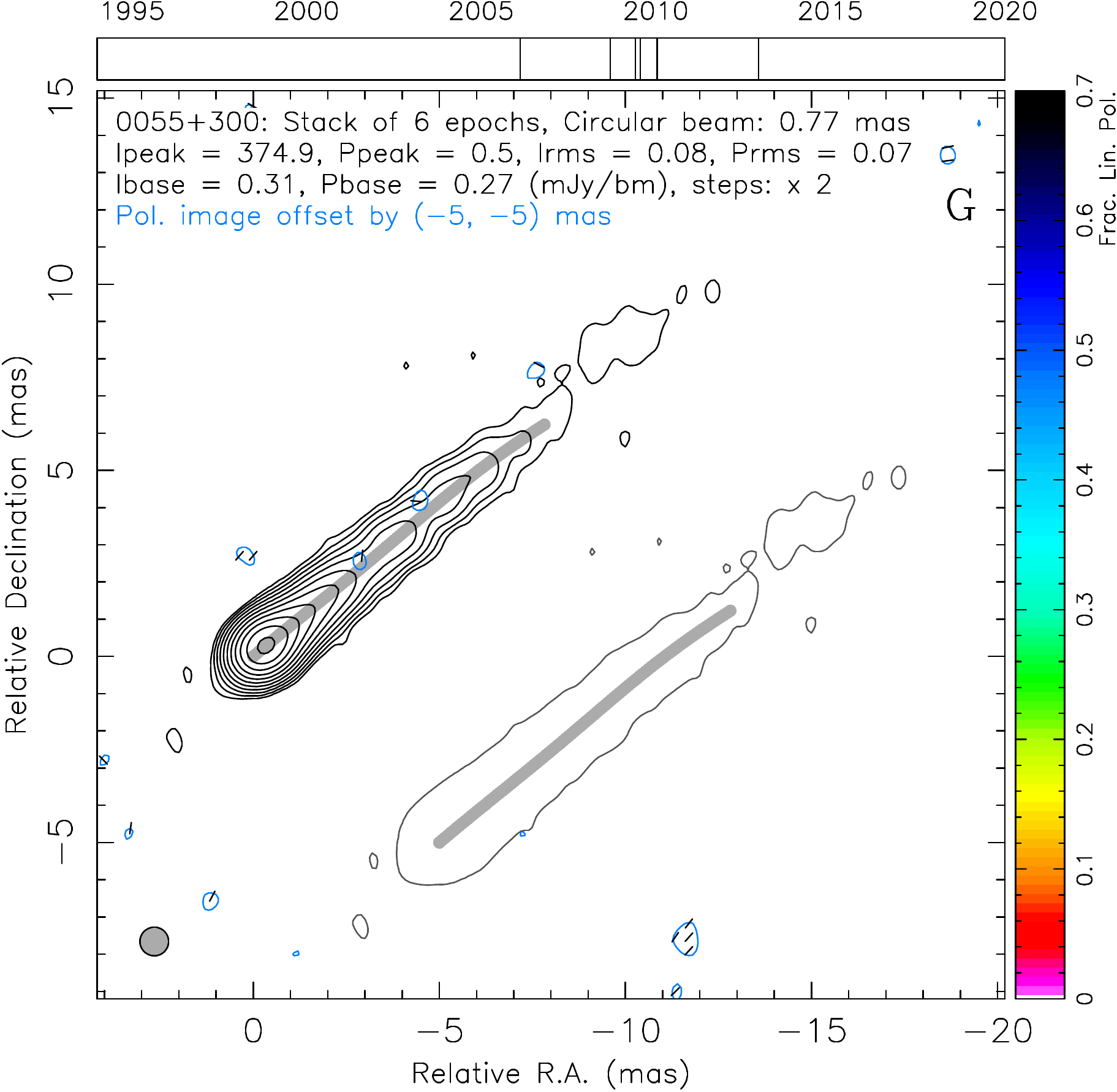}\vspace{0.5cm}
    \caption{Examples of total intensity and linear polarization 15~GHz VLBA stacked CLEAN images of the observed sources with minimum (5, top left), median (9, top right), maximum (139, bottom left) number of epochs and a case of an unpolarized source (bottom right). Each panel contains two contour maps representing (i) total intensity overlaid with a distribution of linear fractional polarization according to the colour wedge, (ii) arbitrarily shifted linearly polarized intensity (blue contours) together with sticks indicating the electric polarization vector directions, not corrected for Faraday rotation. The I and P contours are plotted at increasing powers of 2, starting from the corresponding 4~rms levels. The second map also includes the lowest I contour shown for guidance. The ridgeline constructed in total intensity is shown in gray. The restoring beam is depicted as a shaded circle in the lower left corner. A time wedge indicating the observing epochs (vertical ticks) used for producing the stacked map is shown on top of the image. The source optical class (Q=quasar, B=BL~Lac, G=radio galaxy, U=unknown) is given in the top right corner. 
    (The complete figure set of 436 images is available from Supplementary data online.)
    \label{f:stacked_maps_examples}
    }
\end{figure*}

In \autoref{f:stacked_maps_examples}, we show the examples of the stacked maps having the minimum (5), median (9), maximum (139) number of epochs, and also a source which is unpolarized. For each object, two plots are given: total intensity contours overlaid with a distribution of fractional linear polarization and blue linear polarization contours shifted aside and overlaid with ticks of an arbitrary fixed length representing electric vectors. Their directions have not been corrected for Faraday rotation, which is typically at a level of a few degrees at 15~GHz in the inner jet regions, as found in our earlier rotation measure study of the MOJAVE sample \citep{Hovatta_2012}. We list the parameters of the stacked images in \autoref{t:maps}.

We analyzed the influence of the moving jet components on the stacked maps by comparing (i) a distance they propagate down the jet, $\mu\tau$, where $\mu$ is the maximum proper motion \citep{MOJAVE_XVIII} and $\tau$ is the time interval between the first and last observing epoch and (ii) a typical separation between the components $\Delta r$ derived individually for 398 sources. A median ratio $\mu\tau/\Delta r=1.5$ indicates that the moving components substantially affect the $I$-stacked maps, making them smoother along the outflow. Only about 25 per cent of the stacked maps are not essentially influenced by the moving jet features ($\mu\tau/\Delta r<0.5$), due to either their low proper motion and limited time interval (e.g. in the quasar 
\href{https://www.cv.nrao.edu/MOJAVE/cumulative_stacking_maps/1550+582_stacked_map_evolution.gif}{1550$+$582}) or due to a large separation between the components (e.g. in the radio galaxy 
\href{https://www.cv.nrao.edu/MOJAVE/cumulative_stacking_maps/1345+125_stacked_map_evolution.gif}{1345$+$125}). The influence of the moving jet features on the polarization stacked maps is not that straightforward and might depend on the nature of the bright components. For instance, the jet magnetic field may be turbulent or weakly anisotropic with modest shear causing a net longitudinal field. If the moving components are shocks that compress the magnetic field transverse to the jet \citep{Laing80,Hughes85}, then their polarization will be orthogonal to that of the underlying jet, dominating the apparent polarization if the number and strength of the components is high. If the jet features are plasmoids with random magnetic field, their role would be unimportant. We discuss it further in \autoref{s:EVPA_patterns} based on the inferred EVPA patterns.

\begin{table*}
\centering
\caption{Summary of 15~GHz stacked map parameters.
\label{t:maps}}
\begin{threeparttable}
\begin{tabular}{ccrrcrrcc}
\hline\hline\noalign{\smallskip}
Source &  First Epoch & $\tau$ & N &  Beam & $I_\mathrm{peak}$ & $P_\mathrm{peak}$ & $I_\mathrm{rms}$ & $P_\mathrm{rms}$ \\
       & (yyyy-mm-dd) &    (yr) &   & (mas) &   (mJy bm$^{-1}$) &   (mJy bm$^{-1}$) &  (mJy bm$^{-1}$) &  (mJy bm$^{-1}$) \\
(1) & (2) & (3) & (4) & (5) & (6) & (7) & (8)  & (9) \\
\hline
0003$-$066 & 2003-02-05 &  9.74 &  18 &  0.93 &  1242.8 &  83.0 &  0.11 &  0.05 \\
0003$+$380 & 2006-03-09 &  7.42 &  10 &  0.73 &   462.1 &   3.4 &  0.06 &  0.06 \\
0006$+$061 & 2011-12-29 &  1.43 &   5 &  0.88 &   149.6 &   6.5 &  0.07 &  0.07 \\
0007$+$106 & 2004-02-11 &  9.31 &  13 &  0.86 &   869.3 &   3.0 &  0.05 &  0.05 \\
0010$+$405 & 2006-04-05 &  5.22 &  12 &  0.72 &   508.1 &   3.2 &  0.06 &  0.05 \\
0011$+$189 & 2013-12-15 &  2.22 &   8 &  0.82 &   107.9 &   3.5 &  0.02 &  0.02 \\
0012$+$610 & 2017-01-03 &  2.48 &   6 &  0.66 &   214.2 &   1.0 &  0.04 &  0.03 \\
0014$+$813 & 2016-09-26 &  2.75 &  11 &  0.67 &   381.7 &  17.0 &  0.03 &  0.03 \\
0015$-$054 & 2009-07-05 &  4.01 &   8 &  0.92 &   240.6 &   2.8 &  0.05 &  0.04 \\
0016$+$731 & 2003-08-28 & 15.79 &   8 &  0.66 &  1390.7 &  13.8 &  0.07 &  0.07 \\
\hline
\end{tabular}
\begin{tablenotes}
\item
Columns are as follows:
(1) Source name in B1950.0 coordinates;
(2) Date of first epoch;
(3) Time range between first and last epochs;
(4) Number of single-epoch maps in stack;
(5) FWHM of restoring beam;
(6) $I$-map peak;
(7) $P$-map peak;
(8) $I$-map noise level;
(9) $P$-map noise level. \\
(This table is available in its entirety in a machine-readable form as supplementary material and at the CDS VizieR.)
\end{tablenotes}
\end{threeparttable}
\end{table*}

\subsection{Unpolarized sources}

There are 23 (5.3~per~cent of the sample) essentially unpolarized sources (\autoref{t:unpol_jets}) at the sensitivity level of our stacked images. Eleven of them are radio galaxies, making up 46~per~cent fraction of this source spectral type in the sample. One of these radio galaxies, 0055+300, is shown in \autoref{f:stacked_maps_examples} (bottom right). Four of the radio galaxies, 0128+554, 1404+286, 1509+054 and 2021+614, are classified as compact symmetric objects (CSO). Our sample has one more CSO, i.e. the GPS radio galaxy 1345+125. Polarization emission in this source is detected in the southern jet feature, at about 45~mas from the core \citep{Lister_2003}. However, the integrated fractional polarization over the entire source is relatively low, $m<0.9$ per cent. Unpolarized quasars and BL Lacs are much more rare, both represented by six sources, which makes 2 and 4~per~cent of their fractions, respectively. For one of these quasars, 0710+196, polarization is detected in the core at most of the observing epochs, two of which, 2009 August 19 and 2014 January 25, show the strongest polarization with comparable $P$ flux but roughly orthogonal EVPA directions, thereby canceling each other in vector averaging and resulting in close to zero stacked $P$. Another quasar, 1722+401, demonstrates similar behaviour, with nearly perpendicular EVPAs at the first two and last two epochs. Two BL Lacs, 0329$+$654 and 2013$-$092, are the weakest sources in our sample, making detection of their polarization unlikely due to a sensitivity limit. Taking this into account, only 1~per~cent of the quasars and 3~per~cent of the BL Lacs from our sample are unpolarized. Some sources, e.g. 0615$-$172 and 2021$+$614, show faint and most likely spurious polarization in their stacked $P$-maps due to the $P$-signal being detected at one epoch only, which can be a result of poor leakage term solutions. We excluded these 23 sources from our further analysis.

The reason why some sources of different optical classes, redshifts, apparent speeds and viewing angles are so weakly polarized is unclear. Their polarization sensitivity reached both in single-epoch and stacked images is comparable to that of other AGNs in our sample. It is possible that the weakly polarized sources are subject to strong Faraday depolarization by an external screen associated with an obscuring torus due to large viewing angles of their jets, which is typically the case for radio galaxies \citep[e.g.][]{Aller_2003,Middelberg05}. Internal plasma processes might also play a role, e.g. if a turbulent component of magnetic field in a jet dominates the ordered component on small scales \citep{Marscher_2014}. In this scenario, the chaotic directions of magnetic field effectively cancel out polarization. It is further supported by a significantly higher fraction (almost every second object) of weakly polarized radio galaxies compared to that of blazars, as we probe the outflows on smaller (de-projected) linear separations from the true jet origin, where turbulence is expected to be stronger, and Faraday rotation could be higher because of denser thermal plasma.

\begin{table}
\centering
\caption{Weakly polarized jets.
\label{t:unpol_jets}}
\begin{threeparttable}
\begin{tabular}{lclrrc}
\hline\hline\noalign{\smallskip}
Source & Op. cl. & $z$ &   N & $S_{\rm tot}$ & $P_{\rm tot}$ \\
       &         &     &     &        (mJy)  &         (mJy) \\
   (1) &     (2) & (3) & (4) &           (5) &           (6) \\
\hline
0055$+$300     & G & 0.0165   &  6 &  664.5 & 2.42 \\
0128$+$554$^a$ & G & 0.0365   &  6 &  105.6 & 1.88 \\
0238$-$084     & G & 0.0050   & 24 & 1085.9 & 0.66 \\
1128$-$047     & G & 0.266    & 10 &  698.2 & 1.24 \\
1404$+$286$^a$ & G & 0.077    &  8 &  819.9 & 2.75 \\
1509$+$054$^a$ & G & 0.084    &  8 &  582.3 & 1.75 \\
1637$+$826     & G & 0.024    & 10 &  803.4 & 1.86 \\
1833$+$326     & G & 0.0579   &  8 &  172.1 & 0.92 \\
1845$+$797     & G & 0.0555   & 13 &  439.9 & 1.92 \\
1957$+$405     & G & 0.0561   &  9 & 1437.9 & 2.82 \\
2021$+$614$^a$ & G & 0.227    & 12 & 2212.1 & 3.84 \\
\hline
0646$+$600     & Q & 0.455    &  7 &  742.4 & 2.55 \\
0710$+$196$^b$ & Q & 0.54     & 10 &  222.0 & 2.56 \\
0742$+$103     & Q & 2.624    & 13 & 1354.5 & 1.45 \\ 
1722$+$401$^b$ & Q & 1.049    &  5 &  588.7 & 1.96 \\
2031$+$216     & Q & 0.1735   &  5 &  304.4 & 0.72 \\
2043$+$749     & Q & 0.104    &  7 &  263.0 & 0.90 \\
\hline
0111$+$021     & B & 0.047    &  7 & 505.7 & 1.69 \\
0329$+$654$^c$ & B & \ldots   &  7 &  63.4 & 1.78 \\
0615$-$172     & B & 0.098    &  5 & 159.7 & 0.52 \\
1413$+$135     & B & 0.247    & 12 & 954.3 & 4.11 \\
2013$-$092$^c$ & B & \ldots   &  6 &  52.1 & 1.62 \\
2047$+$098     & B & 0.226    &  6 & 566.5 & 1.80 \\
\hline
\end{tabular}
\begin{tablenotes}
\item
Columns are as follows:
(1) Source name;
(2) Optical class: quasar (Q), BL~Lac (B), radio galaxy (G);
(3) Redshift;
(4) Number of single-epoch maps in stack;
(5) Total cleaned Stokes $I$ flux density in the stacked 15~GHz VLBA image;
(6) Total linear polarization flux density derived from the stacked Stokes $U$ and $Q$ maps as $\sqrt{U^2+Q^2}$.\\
$^a$Compact symmetric object. \\
$^b$Singe-epoch polarization detected in the core. \\
$^c$Weak in total intensity.
\end{tablenotes}
\end{threeparttable}
\end{table}

\subsection{Integrated polarization} \label{subsec:m_int}
\begin{figure}
    \centering
    \includegraphics[width=\linewidth]{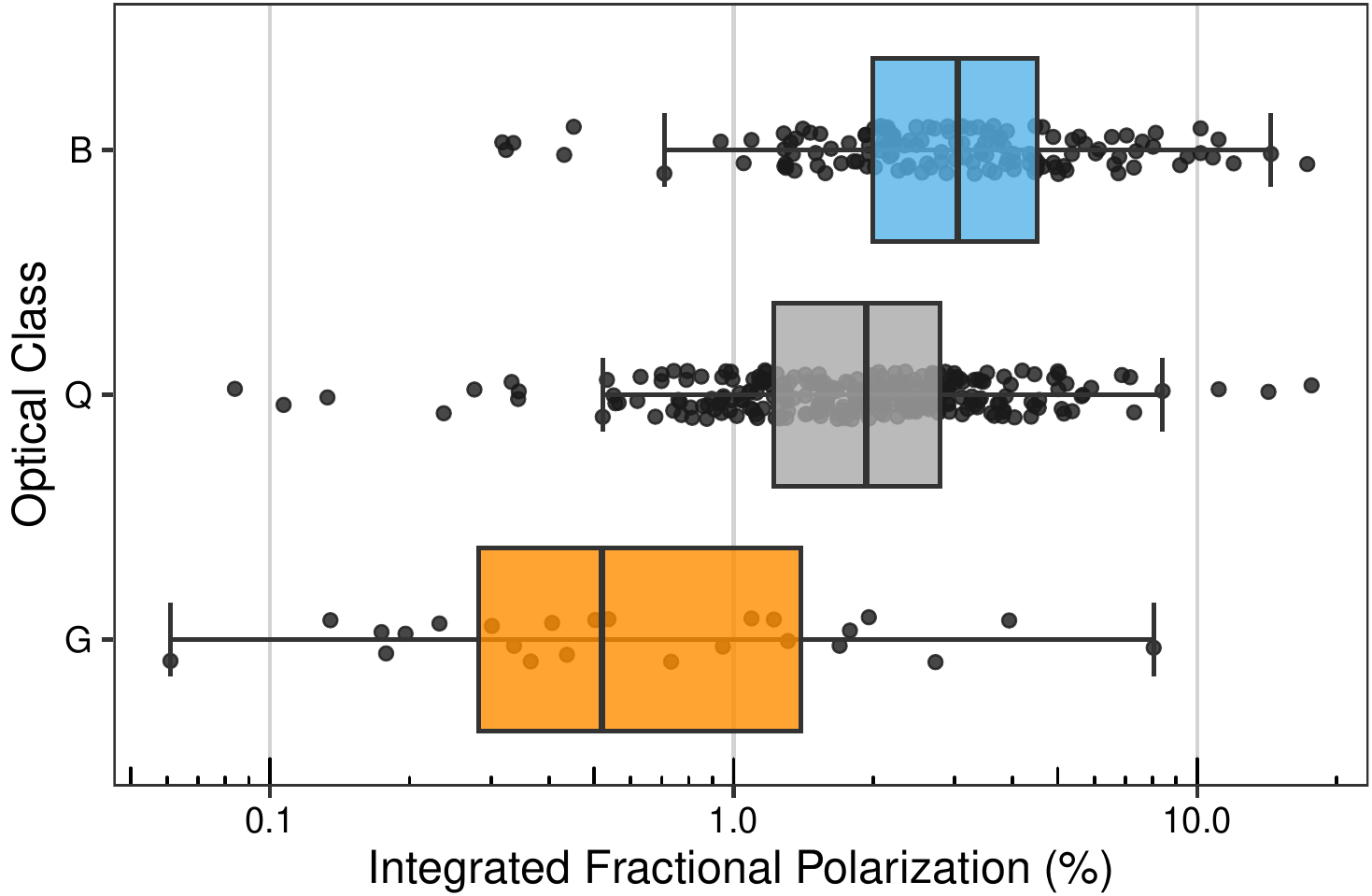}
    \caption{Distributions of linear fractional polarization integrated over the source shown as combined box and scatter plots that break down the distributions by optical class where `B' = BL Lacs (136 objects), `Q' = quasars (259), `G' = radio galaxies (24). The filled regions of the box plots denote the inter-quartile range, while the whiskers extend from the box by 1.5x the inter-quartile range. The vertical lines in the boxes represent medians. Individual data points are shown as a scatter plot over the box plot to better illustrate the range and density of the data.}
    \label{f:m_int}
\end{figure}

\begin{figure}
    \centering
    \includegraphics[width=\linewidth]{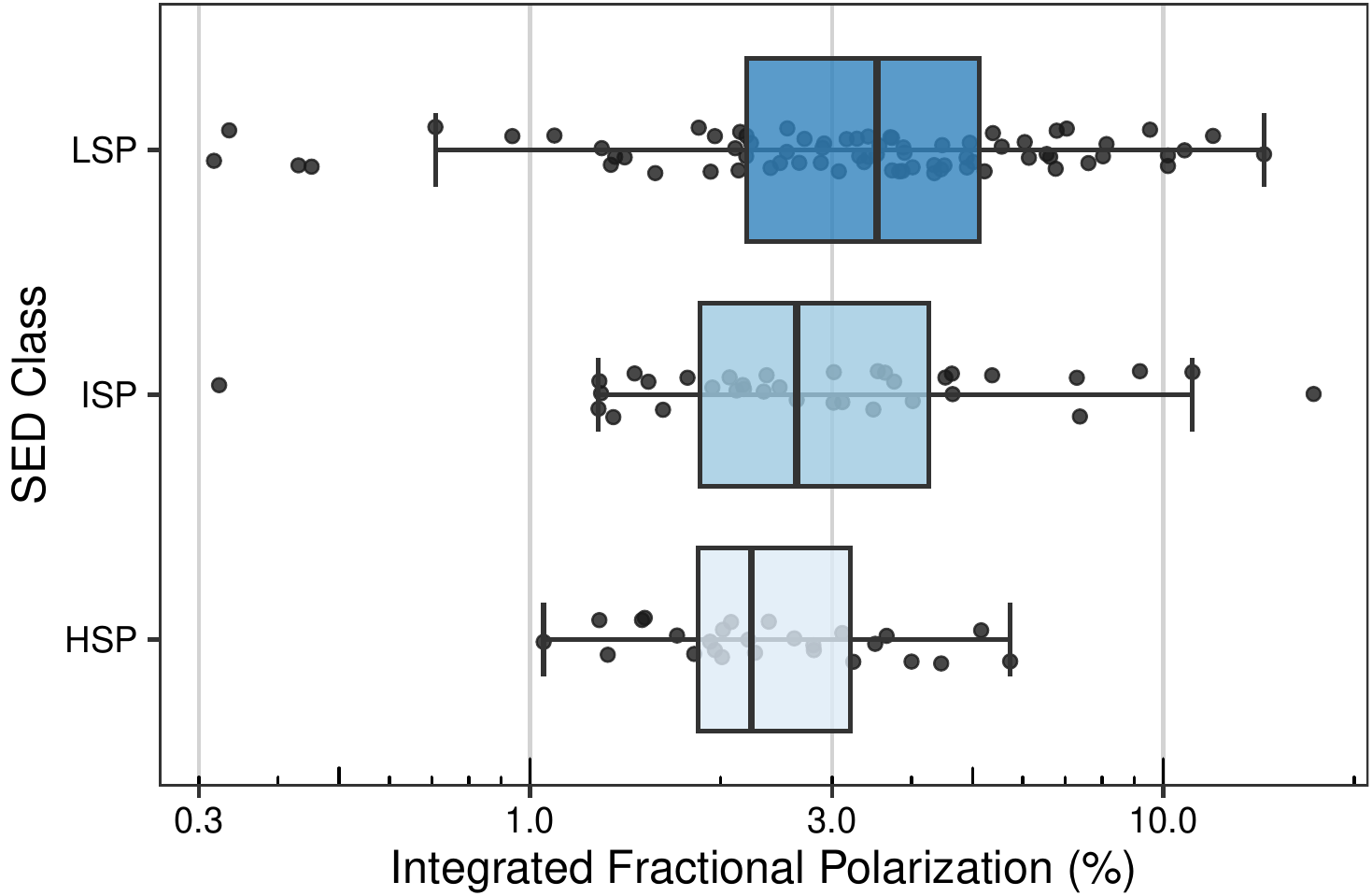}
    \caption{Distributions of linear fractional polarization for the BL Lac objects in our sample as a function of SED class. The `LSP', `ISP' and `HSP' abbreviations indicate low (75 objects), intermediate (35) and high-synchrotron-peak (26) sources, respectively. The scattered points plotted over each box plot indicate the locations of the individual values for a given distribution.}
    \label{f:m_int_bllac_sed}
\end{figure}

We derived the integrated fractional polarization over the entire source as $m_\mathrm{int}=P_\mathrm{tot}/S_\mathrm{tot}$, where $S_\mathrm{tot}$ and $P_\mathrm{tot}$ are the total flux density and polarized flux, respectively, calculated over the image above the corresponding $4I_\mathrm{rms}$ and $4P_\mathrm{rms}$ levels. The values of $m_\mathrm{int}$ are distributed in a wide range, from 0.06~per cent for the galaxy NGC\,1052 to 17.6~per cent for the quasar 3C\,119. The distributions of $m_\mathrm{int}$ for the sources of different optical classes are shown in \autoref{f:m_int}. The least polarized sources are radio galaxies, with a median $m_\mathrm{int}=0.52\pm0.24$~per cent. According to a Kolmogorov-Smirnov test, quasars and BL Lacs have statistically different distributions, with the medians of $m_\mathrm{int}$ equal to $1.93\pm0.09$ and $3.05\pm0.24$~per cent, respectively. BL Lacs being devided into different SED classes show that HSP sources are on average less polarized than ISP and LSP BL Lacs (\autoref{f:m_int_bllac_sed}).

\subsection{Total intensity ridgelines} \label{subsec:ridgelines}
We traced out the total intensity ridgeline for each stacked map, following the approach described in \citet{Pushkarev_2017}. It is based on making an azimuthal slice around the core component at a given radius and finding a position of the weighted average in total intensity. The starting ridgeline point coincides with the core position and then the algorithm advances downstream for successively increasing core separations with a step of 0.5~mas, skipping the regions where emission is too faint, $<4I_\mathrm{rms}$. After that we applied a cubic spline interpolation to have a ridgeline point approximately every 0.1~mas along the ridgeline. In the cases when the counter-jet is detected, i.e. in the radio galaxies TXS\,0128+554, NGC\,1052, 3C\,84, PMN\,J1511+0518, Cygnus A, quasar S4\,0646+60 and BL Lac PKS\,B1413+135, we show the ridgeline for the approaching, brighter jet.

\section{Polarization properties along and across jets} \label{sec:p_prop}

\subsection{Fractional polarization}
\label{s:m_vs_r}

The first evidence that the degree of linear polarization of distinct jet features increases with their separation from the core was reported by \citet{Cawthorne93}, who analysed polarimetric VLBI data at 5~GHz for a small sample of AGN jets. Later, this trend  was also detected at 22 and 43~GHz \citep{ListerSmith00,Lister01} for samples of 18 and 32 sources, respectively.  At 15~GHz, the trend was found by \citet{ListerHoman05}, who studied the first-epoch MOJAVE data for the flux-density-limited complete sample of 133 sources. This observational result implies that the magnetic field, decreasing with distance from the true jet vertex as $B\propto r^{-k}$ ($k=2$ for the case of the poloidal field and $k=1$ for the toroidal field and a conical jet shape), becomes more ordered, if it is not entirely caused by, e.g. decreasing depolarization along the jet.

\begin{figure}
    \centering
    \includegraphics[width=\linewidth]{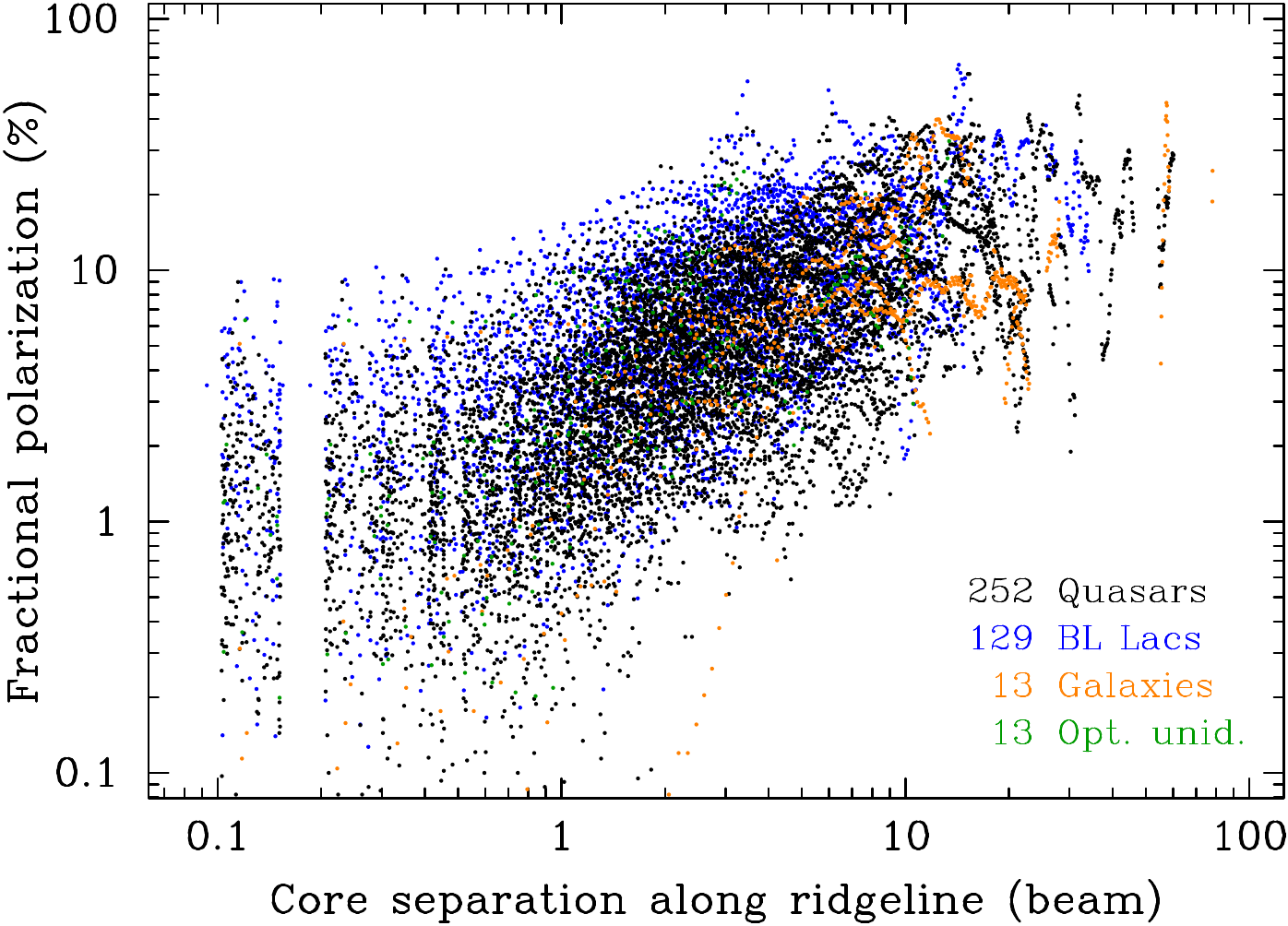}
    \caption{Linear fractional polarization vs projected distance from the core along the ridgeline measured in beam sizes. 
    \label{f:m_vs_r}
    }
\end{figure}

In \autoref{f:m_vs_r}, we show the dependence of fractional polarization as a function of the projected, i.e. observed, distance down the jet along its ridgeline for the largest ever sample of over 400 sources. We confirm the general increase in $m$ downstream and discuss its features. First, the degree of linear polarization in the core region (within $\sim$0.5~beam) ranges from about 0.1~per~cent to 10~per~cent, with a median of about 1~per~cent and 2~per~cent for quasars and BL Lacs, respectively, whereas the radio galaxy cores are found to be low or completely unpolarized. The synchrotron emission of the core is partially opaque, leading to a further decrease in the fractional polarization both through opacity and possibly increased Faraday effects if they are present. The low $m_\mathrm{core}$ values are also caused by in-beam depolarization in the case of superposition from polarized emission of unresolved features with different EVPAs. 

Beyond the core in the inner jet, $m$ increases quite rapidly, mostly due to a switch to an optically thin regime of synchrotron radiation and weakening of the beam depolarization effect. In the outer jet regions, the degree of polarization, on average, continues to grow with distance from the core. This can be explained by weakening shocks and accompanied turbulence of emitting plasma \citep{Marscher_2014}, progressively smaller Faraday rotation \citep{Hovatta_2012,2017MNRAS.467...83K} and the spectral ageing effect \citep{Kardashev62}. In the latter phenomenon, the high-energy electrons lose energy faster owing to synchrotron emission. This leads to steepening of both the energy and flux density spectra in the more remote jet parts. VLBI observations showed that the magnitude of the spectral ageing effect is about $\Delta\alpha\simeq -0.6$ for parsec-scale AGN jets \citep{PK12_RDV,Hovatta_2014}. In turn, this results in the growth of $\Delta m_\mathrm{jet}\sim0.1$ because for the optically thin regime of synchrotron radiation from a region with an ordered magnetic field and randomly oriented pitch angles of relativistic electrons the corresponding maximum of $m$ is limited by $(p+1)/(p+7/3)$ \citep{Pacholczyk70}, where $p=1-2\alpha$ is the power-law index in the energy spectrum of emitting particles $N(E) \propto E^{-p}$.

\begin{figure}
    \centering
    \includegraphics[width=\linewidth]{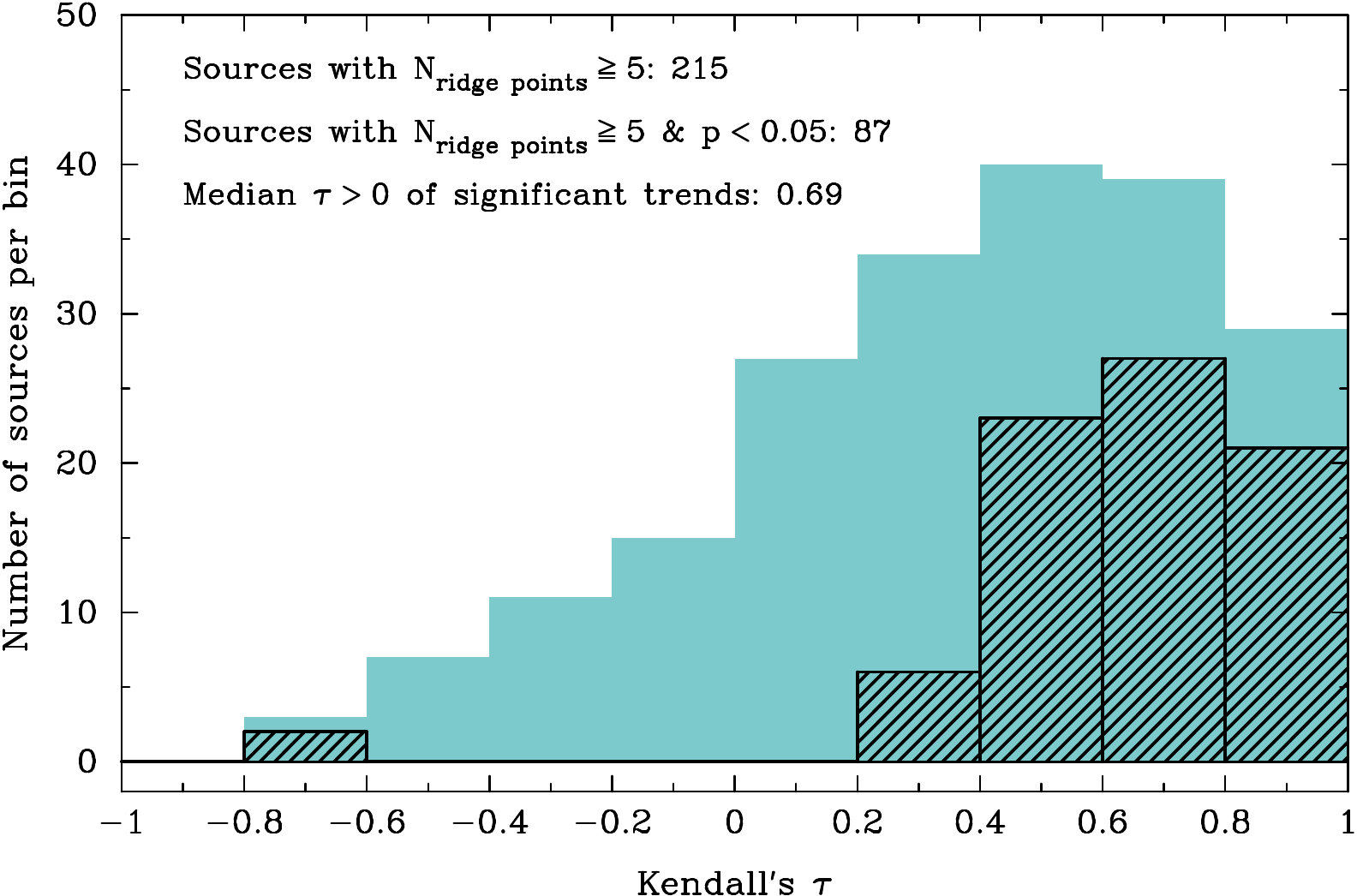}
    \caption{Histogram of the Kendall's $\tau$ correlation coefficient between $m$ and the distance from the core along the ridgeline, derived individually for the sources having at least five $P$-sensitive ridgeline points taken from a single pixel every half beam size beyond the one-beam core region. The hatched bins represent sources with a significant (chance probability $p<0.05$) correlation.
    \label{f:m_vs_r_hist}
    }
\end{figure}

The growth of fractional polarization down the jet can also occur either through shear in a boundary layer \citep[e.g.][]{Laing14} or as the pitch of a helical magnetic field decreases with distance from the origin \citep{Porth11}. As shown by \cite{BP22}, the $m$-values on the jet axis gradually increase with decreasing the pitch angle, starting from about $55^\circ$. This was derived from modelling transverse slices of fractional polarization for a wide space of geometric and kinematic jet parameters, e.g. the viewing angle ranging from $2^\circ$ to $10^\circ$ in the observer's frame and Doppler factors varying from 4 to about 15. This mechanism, likely contributing insignificantly in the innermost jet parts, can progressively add to the slow logarithmic increase in $m$ and ultimately dominate on large scales. The continuous growth in fractional polarization with the core separation, together with close alignment of EVPA with the local jet direction (typically seen in BL Lacs, more in \autoref{s:EVPA_patterns}), allows us to argue for a large-scale ordered component, i.e. helical magnetic field rather than the presence of an isolated jet feature with high fractional polarization \citep[e.g.][]{Pushkarev00}.

We also analysed individual dependencies of $m$ along the ridgelines. To get rid of a neighbouring-pixel dependence in a map arising due to convolution with the restoring beam, we interpolated the cubic spline of the ridgeline with a step of half a beam size. For the sources which had five or more such ridgeline points with $P>4\sigma_P$ beyond the core region, we calculated the Kendall's $\tau$ correlation coefficient (\autoref{f:m_vs_r_hist}). In total, 215 sources satisfied these criteria, 87 of which show significant (with random chance probability $p<0.05$, hatched area) positive correlation, while only two objects, 0509+406 and 0603+476, have negative correlation.

\begin{figure}
    \centering
    \includegraphics[width=\linewidth]{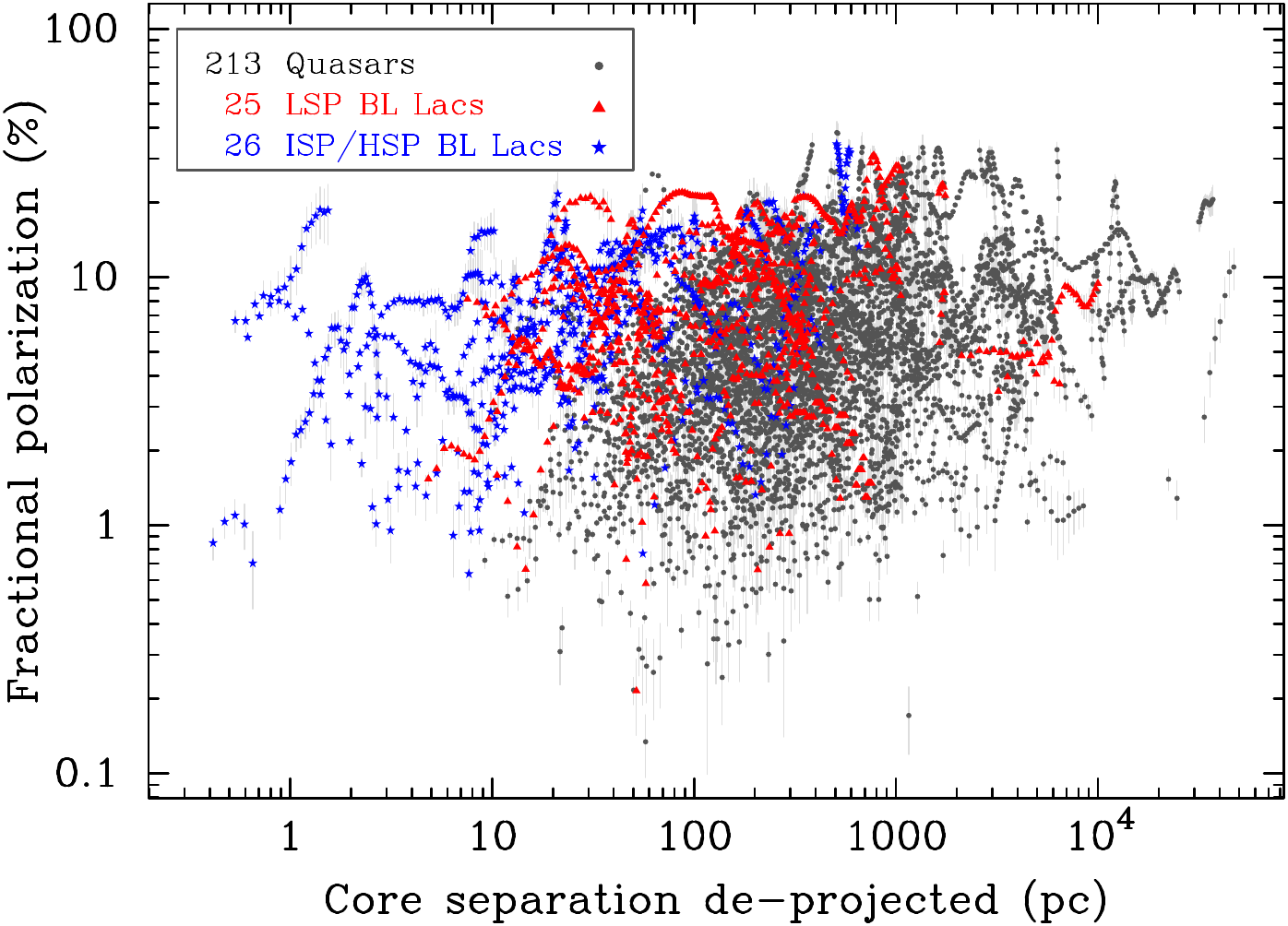}
    \caption{Linear fractional polarization of quasars and BL Lacs against the de-projected distance from the core along the ridgeline starting beyond the one-beam core area.
    \label{f:m_vs_r_BQ}
    }
\end{figure}

As seen in \autoref{f:m_vs_r}, the ridgeline $m$-values of BL Lacs at a given projected distance from the core are typically higher than those of quasars. This confirms earlier results obtained by \citet{Lister01} on a smaller source sample. To investigate it further, we first converted the angular distances to a linear projected scale for the sources with known redshifts and then corrected for projection by applying the viewing angles derived from the estimates of Doppler factors \citep{MOJAVE_XIX} and apparent jet speeds \citep{MOJAVE_XVIII}. The ridgeline $m$-values beyond the one-beam core region of 52 BL Lacs and 211 quasars vs de-projected distance in parsecs are given in \autoref{f:m_vs_r_BQ}. It shows that the jet-axis fractional polarization of BL Lacs probed on smaller scales (up to decaparsecs) are considerably higher than those in quasars, but the corresponding slope is flatter, thus intersecting the steeper quasar trend on hectoparsec scales, showing comparable $m$-values. This can be explained by higher rotation measures \citep{Hovatta_2012} or initially larger helical B-field pitch angle or higher rates of energy losses in quasars compared to those of BL Lacertae objects. The HSP (17), ISP (9) and LSP (25) BL Lacs are characterised by on average decreasing viewing angles as found by \cite{MOJAVE_XIX} and also increasing redshifts. They form different partly overlaid segments of the whole $m$-trend, showing generally higher fractional polarization of LSP BL Lacs. If we compare quasars and LSP BL Lacs which have matched redshifts $z\lesssim0.5$, there is no systematic difference in fractional polarization along their jet ridgelines.

\begin{figure}
    \centering
    \includegraphics[width=\linewidth]{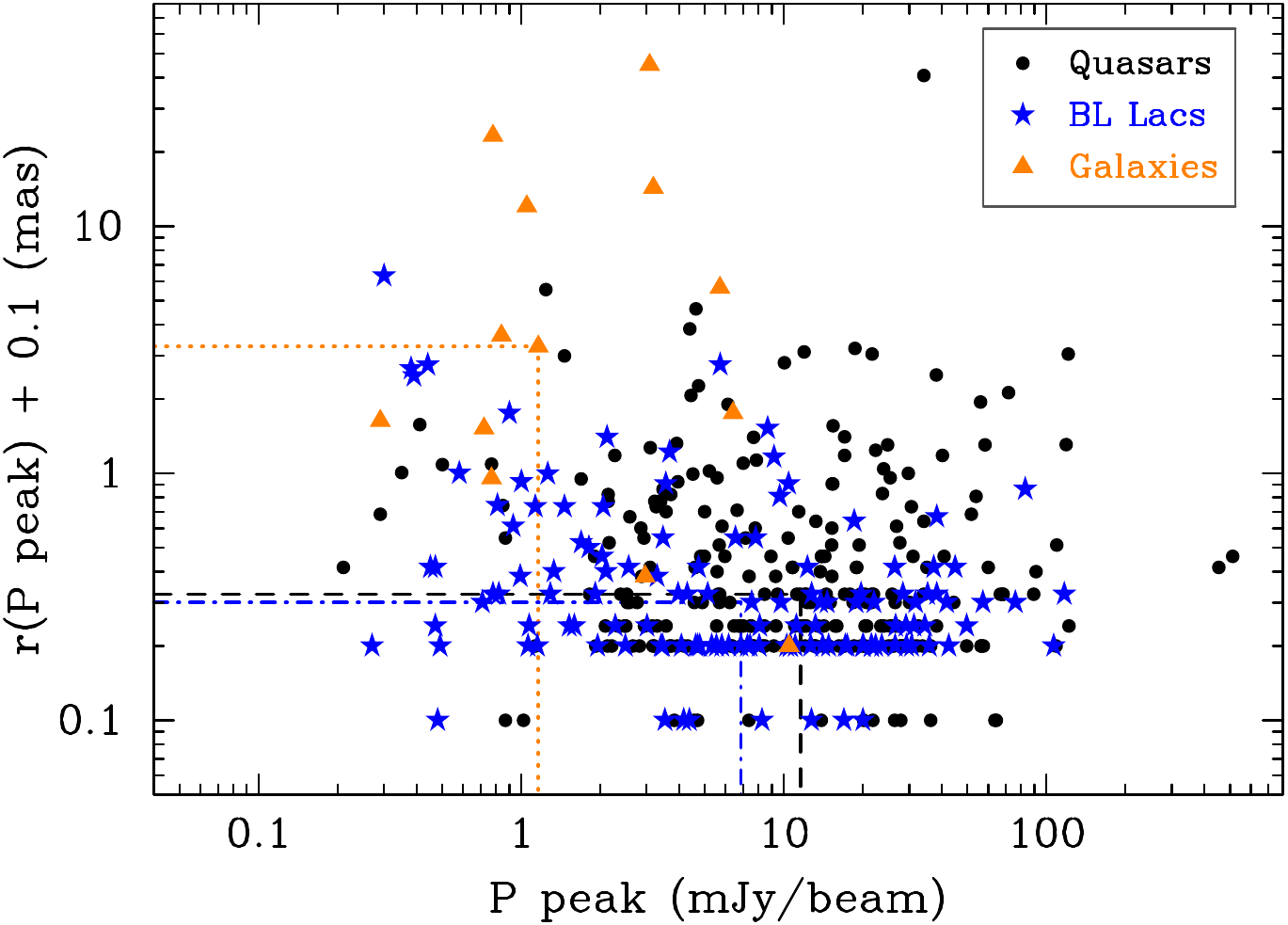}
    \caption{Angular separation between the position of the polarization intensity peak and the total intensity core vs $P_\mathrm{peak}$ for 253 quasars, 129 BL Lacs and 13 radio galaxies. The dashed lines indicate the medians for each optical class.
    \label{f:p_peak_offset}
    }
\end{figure}

\begin{figure*}
    \centering
    \includegraphics[height=0.32\linewidth]{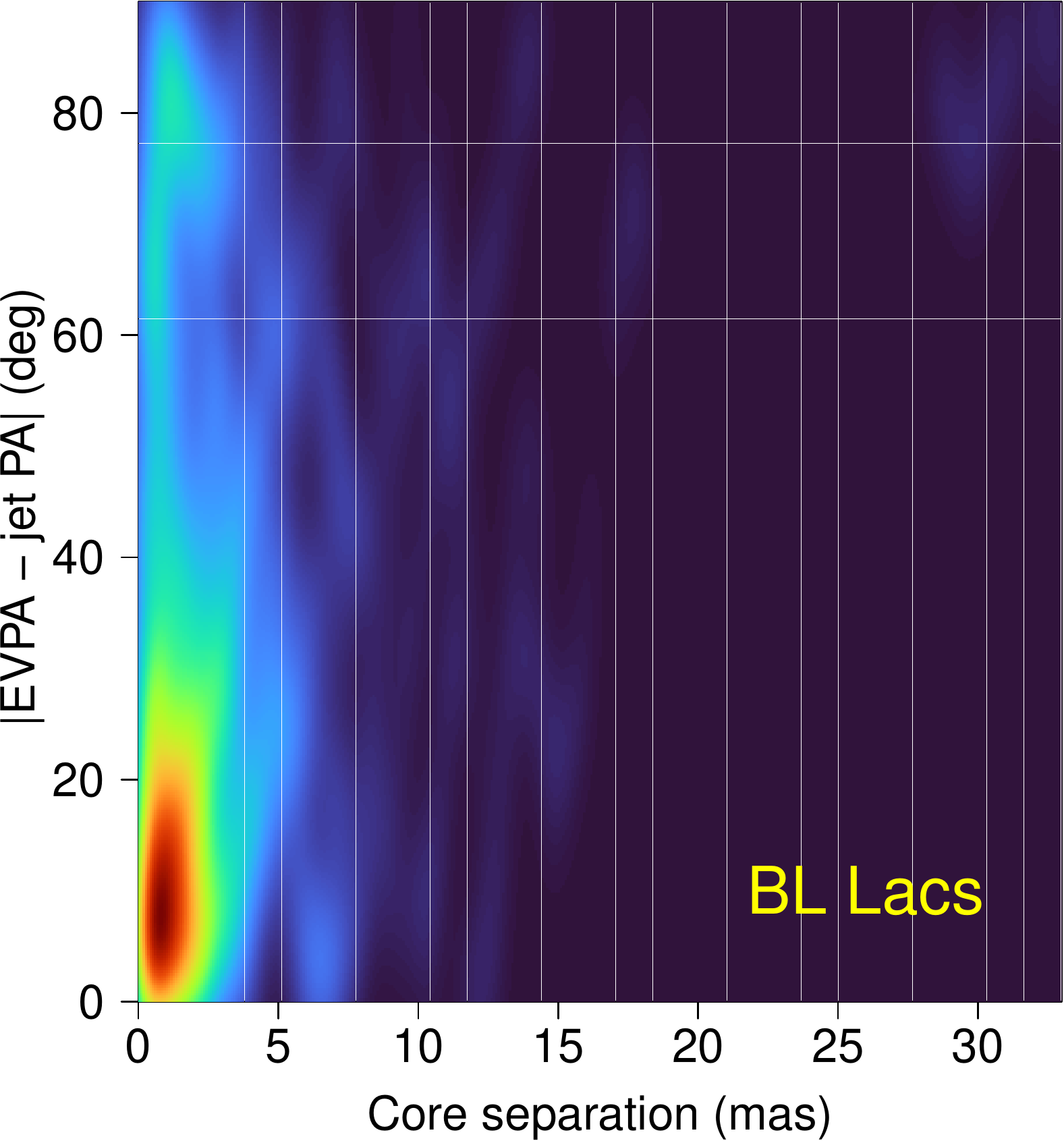}\hspace{0.1cm}
    \includegraphics[height=0.32\linewidth]{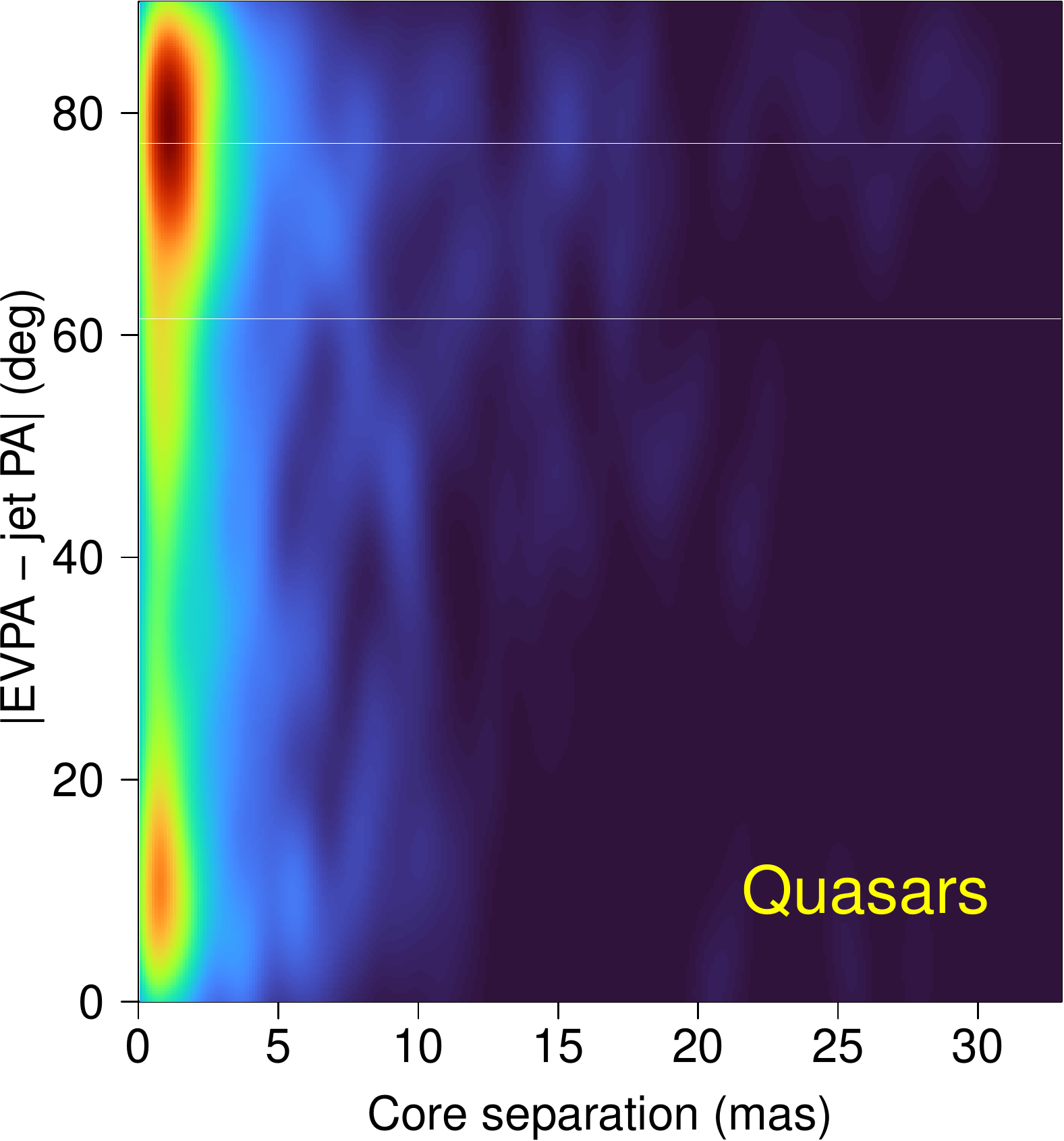}\hspace{0.1cm}
    \includegraphics[height=0.32\linewidth]{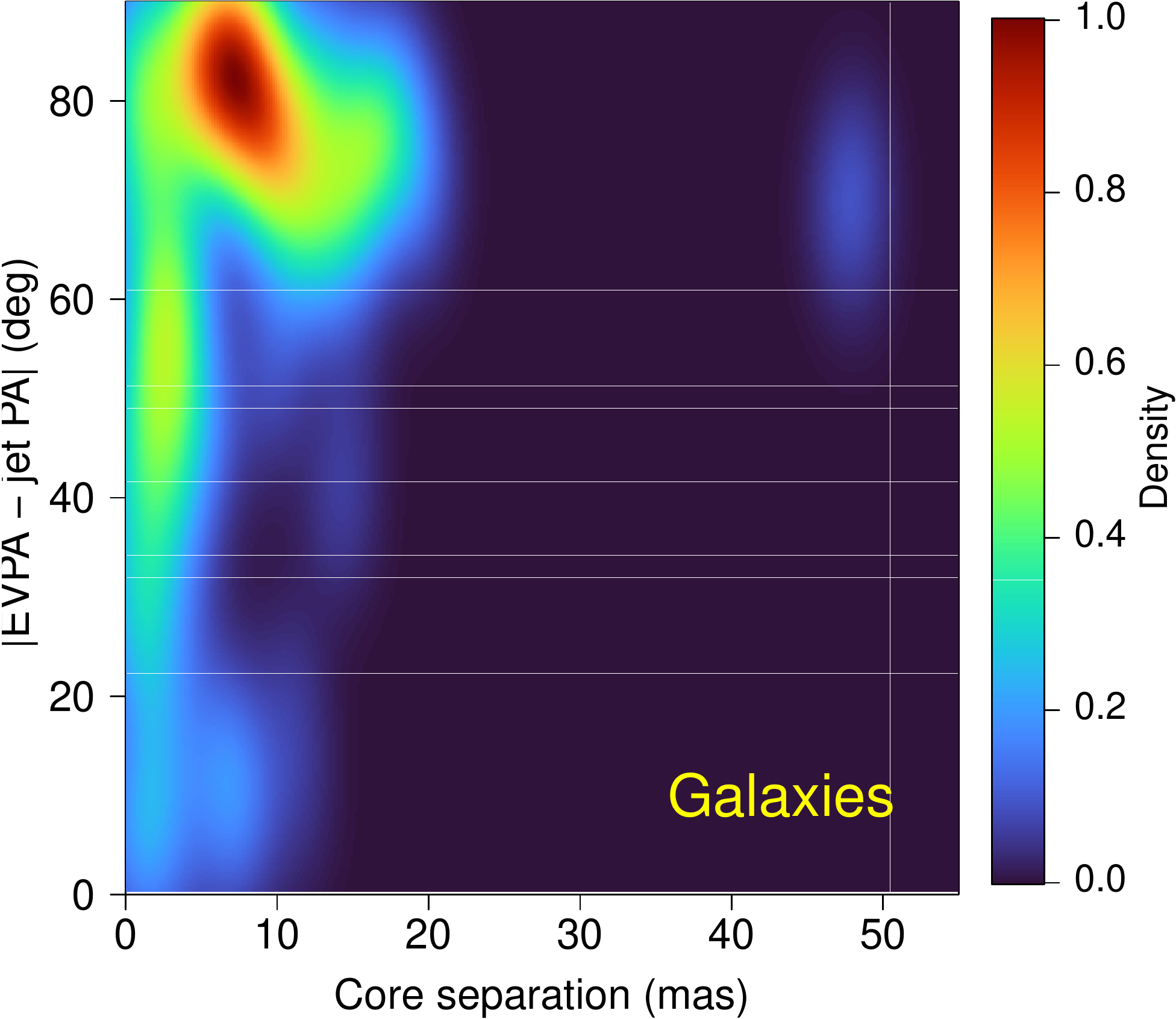}
    \caption{Density distribution of the absolute deviation of EVPA from the local jet direction vs the core separation along the ridgeline for 129 BL Lacs (left), 252 quasars (middle) and 12 radio galaxies (right).
    \label{f:evpa_jetpa_offset}
    }
\end{figure*}

\subsection{Polarization intensity peak position}
The brightness distribution of AGN jets on parsec scales in total intensity is typically dominated by a bright core for the sources of all optical classes. It is not the case for linearly polarized emission. Thus, the cores and innermost jet regions of radio galaxies are often unpolarized, and the median $P$-peak position is about 3~mas away from the apparent jet origin (\autoref{f:p_peak_offset}). BL Lacs and quasars have their polarization intensity peaks much closer to the core, at a median distance of about 0.2~mas downstream the jet. The non-zero offset is caused by opacity effects in the core region, so that polarization is dominated by the innermost jet components, the emission of which becomes optically thin. The $P$-peak position coincides with the core within the pixel size of 0.1~mas in only $\sim$6~per~cent of cases for quasars and BL Lacs and none for radio galaxies. 

Quasars are found to be strongest in polarization intensity, while BL Lacs have an almost two times weaker median $P$ peak, and radio galaxies are about an order of magnitude weaker. However, we remind that in terms of fractional polarization, the BL Lacs are stronger in the core. There are a few exceptional cases when quasars, like radio galaxies, do not reveal any polarization in the core but show a high SNR $P$ emission in another jet part. These are 
\href{https://www.cv.nrao.edu/MOJAVE/cumulative_stacking_maps/1602+576_stacked_map_evolution.gif}{1602+576} 
with the only polarized feature at about 5~mas from the core, 
\href{https://www.cv.nrao.edu/MOJAVE/cumulative_stacking_maps/0821+621_stacked_map_evolution.gif}{0821+621} 
and 
\href{https://www.cv.nrao.edu/MOJAVE/cumulative_stacking_maps/1435+638_stacked_map_evolution.gif}{1435+638} 
with detected polarization beyond the core, at separations of about 1~mas. Interestingly, no BL Lacs show such a phenomenon.

\subsection{EVPA and jet direction}
Earlier, studying the core polarization properties, we found that the EVPAs in the VLBI cores of BL Lacs are better aligned with the innermost jet direction and less variable than in quasars \citep{MOJAVE_XVI}. Here, we analyse the EVPA orientation with respect to the local jet direction along the ridgeline. In \autoref{f:evpa_jetpa_offset}, we plot two-dimensional histograms of $|\mathrm{EVPA} - \mathrm{jetPA}|$ vs core distance separately for different optical classes of the sources. BL Lacs do show a tendency for EVPA to follow the jet. It is clearly seen in about 60 per cent of BL Lacs. This confirms earlier results taken by \cite{Gabuzda00} on the polarized core and jet components from a sample of 24 BL Lacertae objects. The offset of the density peak of the unsigned EVPA deviations from zero is likely caused by a limited accuracy of the local jet direction from ridgelines and noise error on the EVPAs, rather being a real physical effect. Indeed, if the sign of EVPA deviation is taken into account the density peak spreads around zero nearly symmetrically. Only 14 objects (10 per cent) currently classified as BL Lacs, namely,
\href{https://www.cv.nrao.edu/MOJAVE/cumulative_stacking_maps/0141+268_stacked_map_evolution.gif}{0141$+$268}, 
\href{https://www.cv.nrao.edu/MOJAVE/cumulative_stacking_maps/0214+083_stacked_map_evolution.gif}{0214$+$083}, 
\href{https://www.cv.nrao.edu/MOJAVE/cumulative_stacking_maps/0313+411_stacked_map_evolution.gif}{0313$+$411}, 
\href{https://www.cv.nrao.edu/MOJAVE/cumulative_stacking_maps/0518+211_stacked_map_evolution.gif}{0518$+$211}, 
\href{https://www.cv.nrao.edu/MOJAVE/cumulative_stacking_maps/0723-008_stacked_map_evolution.gif}{0723$-$008}, 
\href{https://www.cv.nrao.edu/MOJAVE/cumulative_stacking_maps/0735+178_stacked_map_evolution.gif}{0735$+$178}, 
\href{https://www.cv.nrao.edu/MOJAVE/cumulative_stacking_maps/0845-068_stacked_map_evolution.gif}{0845$-$068}, 
\href{https://www.cv.nrao.edu/MOJAVE/cumulative_stacking_maps/0851+202_stacked_map_evolution.gif}{0851$+$202}, 
\href{https://www.cv.nrao.edu/MOJAVE/cumulative_stacking_maps/1011+496_stacked_map_evolution.gif}{1011$+$496}, 
\href{https://www.cv.nrao.edu/MOJAVE/cumulative_stacking_maps/1133+704_stacked_map_evolution.gif}{1133$+$704}, 
\href{https://www.cv.nrao.edu/MOJAVE/cumulative_stacking_maps/1215+303_stacked_map_evolution.gif}{1215$+$303}, 
\href{https://www.cv.nrao.edu/MOJAVE/cumulative_stacking_maps/1219+285_stacked_map_evolution.gif}{1219$+$285}, 
\href{https://www.cv.nrao.edu/MOJAVE/cumulative_stacking_maps/1514-241_stacked_map_evolution.gif}{1514$-$241}, 
\href{https://www.cv.nrao.edu/MOJAVE/cumulative_stacking_maps/2010+463_stacked_map_evolution.gif}{2010$+$463}, 
manifest EVPA roughly orthogonal to the jet direction. If we divide BL Lacs into SED classes, the sources with the aligned EVPA are fractionally distributed very much close to the whole BL Lac sub-sample (55 per cent of LSP, 26 per cent of ISP, and 26 per cent of HSP). However, those characterised by the EVPA transverse to the local jet position angle have a roughly double relative excess of HSP and the corresponding lack of ISP sources. About 26 per cent of BL Lacs show either non-monotonic orientation of EVPA along the jet or reveal polarization in the core region only, with EVPA not aligned with the jet. Some of these sources might represent EVPA aligned with the innermost jet direction if the outflow is bent there but not resolved. The remaining 4 per cent (six sources) are unpolarized objects, listed in \autoref{t:unpol_jets}.

The $|\mathrm{EVPA}-\mathrm{jetPA}|$ in quasars show all possible values, with a clear predominance of a peak near $90^\circ$ and a weaker secondary peak at values close to $0^\circ$, in contrast to BL Lacs. There is also an indication that EVPAs in quasars and radio galaxies become preferentially transverse to the jet in the outer ridgeline regions. This analysis suggests that $|\mathrm{EVPA} - \mathrm{jetPA}|$ might be a good discriminator between flat-spectrum radio quasars and BL Lacertae objects.

\subsection{Cumulative stacking}
\begin{figure}
    \centering
    \includegraphics[width=\linewidth]{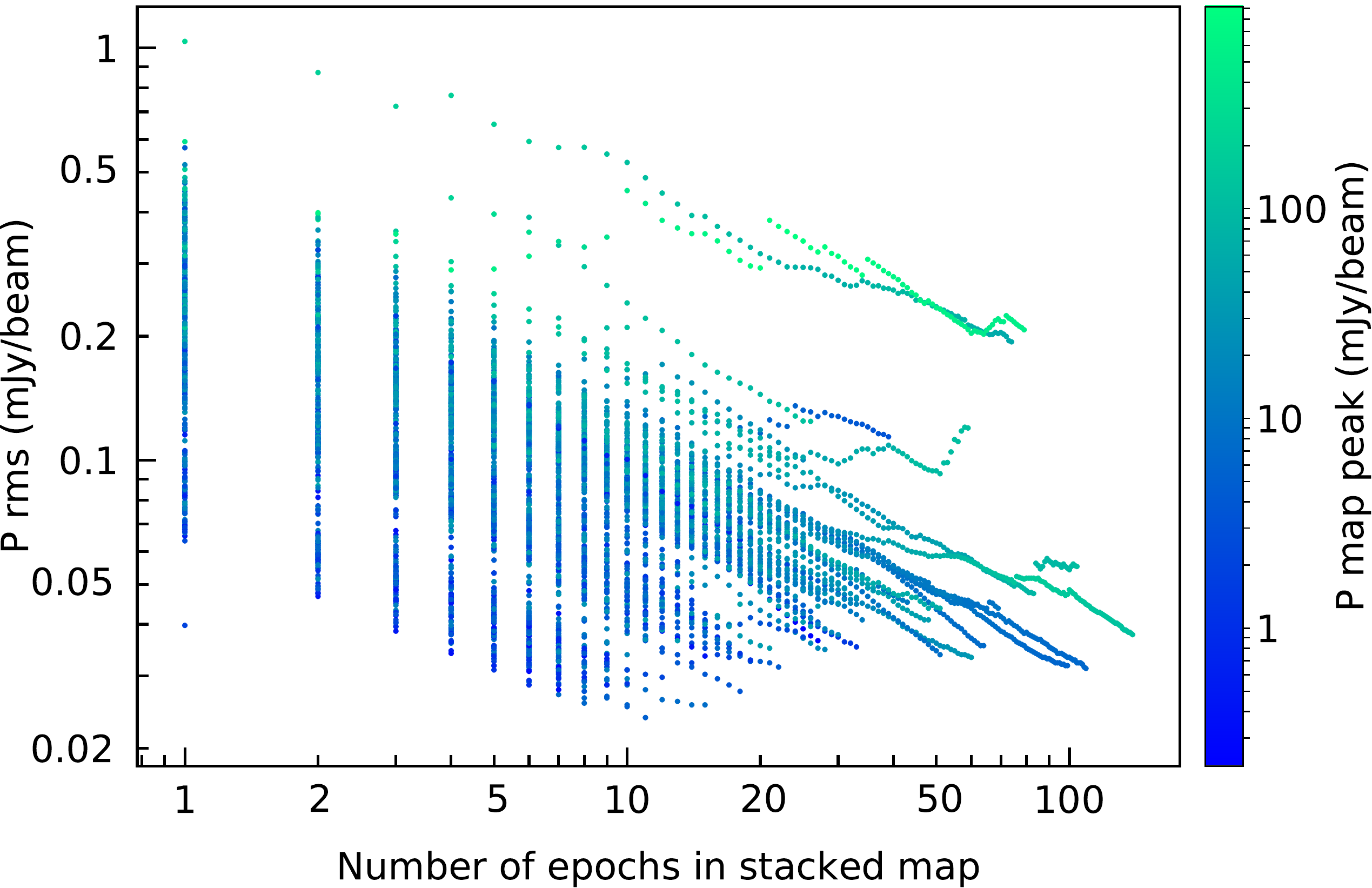}
    \includegraphics[width=\linewidth]{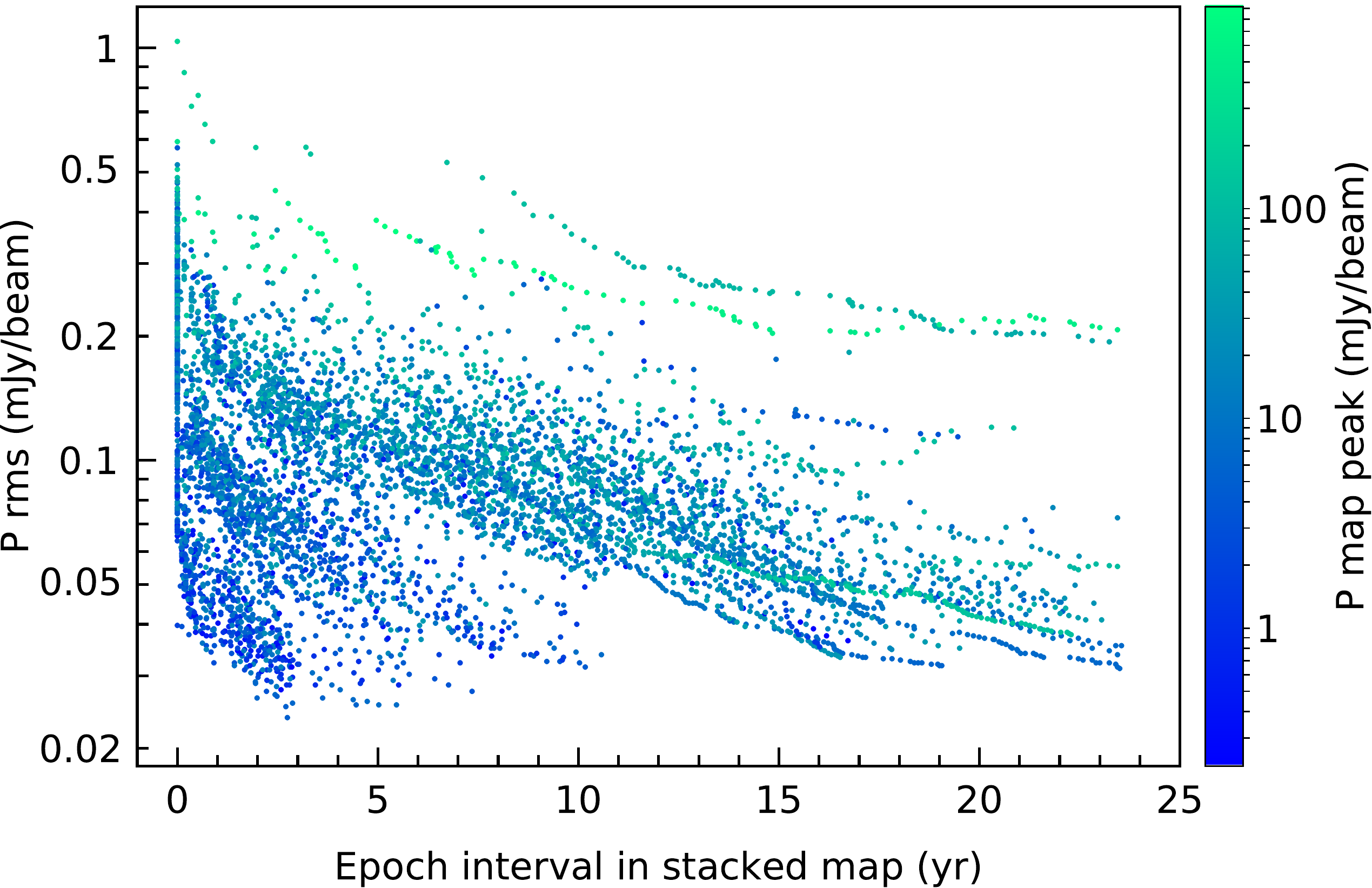}
    \caption{Noise level of the cumulative $P$ stacked maps vs the number of epochs (top) and vs the time interval between the first and last epochs in a stack (bottom).
    \label{f:p_rms} 
    }
\end{figure}

We also produced a series of all intermediate stacked maps for every source by successively adding all available later epochs to the first one to analyse the changes in the image parameters. As expected, the noise level both in $I$ and $P$ reduces the following $N_\mathrm{epoch}^{-0.5}$ dependency (\autoref{f:p_rms}, top). The evolution of $\sigma_P$ against the time interval of epochs in a stacked map is shown in \autoref{f:p_rms} (bottom). The noise reaches values down to about 30~$\mu$Jy~beam$^{-1}$. The three stripes from upper to mid and lower reflect observations carried out with initially different bit-rates of 128 (until mid-2007, brighter sources), 512 and 2048~Mb~s$^{-1}$ (since 2014, weaker sources). Stacking reduced the rms noise by a factor of a few for most cases and up to ten times for the most frequently observed sources. The corresponding dependencies for $\sigma_I$ look qualitatively similar.

\begin{figure}
    \centering
    \includegraphics[width=\linewidth]{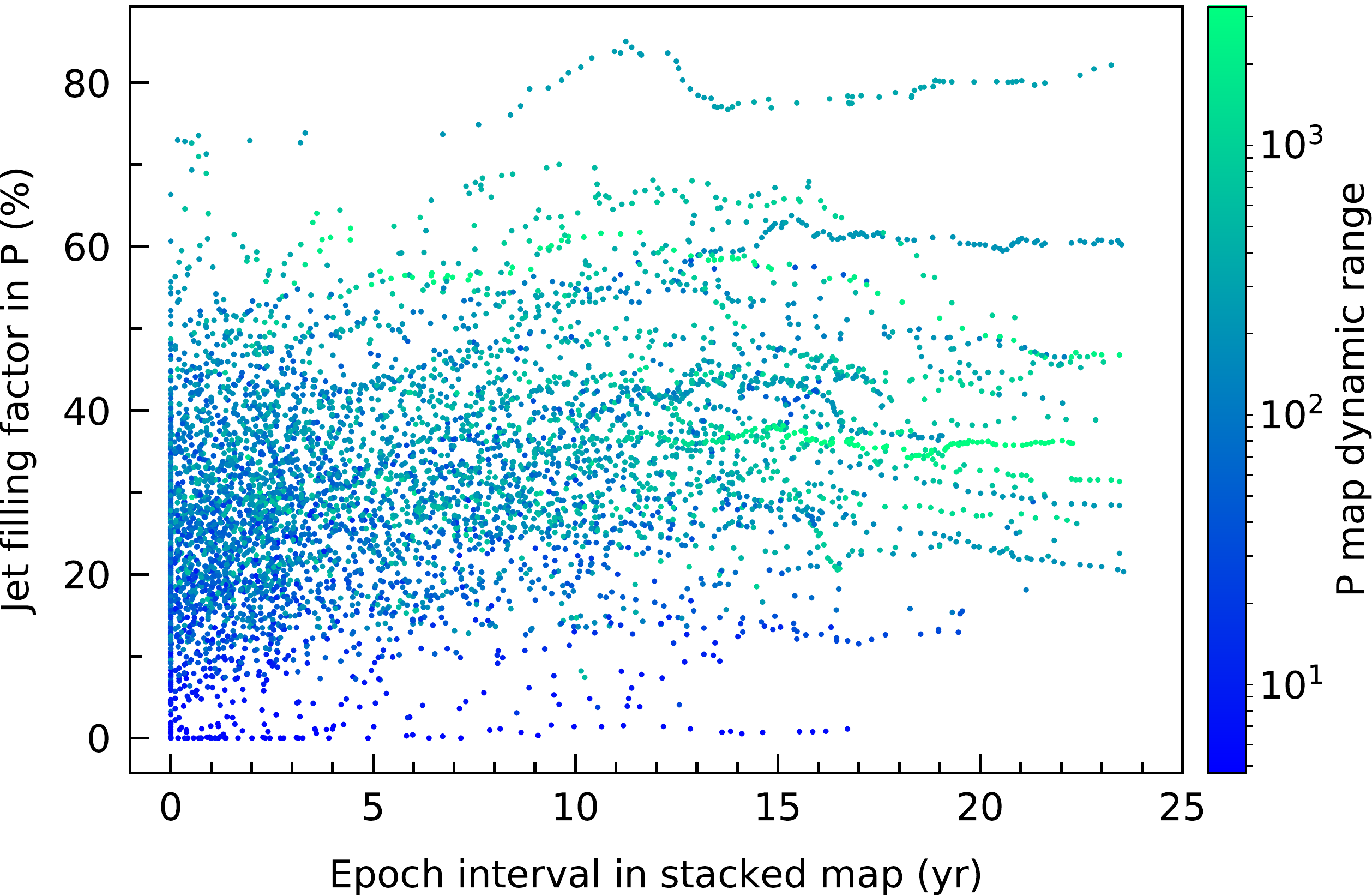}
    \caption{Fraction of the $P$-sensitive source area with respect to that of $I$ as a function of a time interval between the first and last epochs in stacked maps, produced by cumulatively adding epochs. The colour bar denotes the dynamic range of a stacked $P$ map derived as $P_\mathrm{peak}/\sigma_P$.
    \label{f:p_satur}
    }
\end{figure}

In our earlier study of jet shapes and their opening angles \citep{Pushkarev_2017}, we found that the apparent jet opening angle increases if more epochs are added to the stacked map. This growth is not monotonic, it plateaus on scales of about five years, implying that stacking revealed a full cross-section of the inner jet. In the current paper, we analyse the effect of stacking on linearly polarized emission. Combining data of different epochs results in higher image sensitivity, enlarging the areas of the revealed jet structure in total and linearly polarized intensity, $S_I$ and $S_P$, where $S_I$ is the area within the bottom ($4\sigma_I$) total intensity contour and $S_P$ is the area within an overlap of the bottom linearly polarized ($4\sigma_P$) and total intensity ($4\sigma_I$) contours. Thus, the image $P$-filling factor of a jet defined as $S_P/S_I$ is limited by 1. In addition to the number of epochs in a stack, their temporal spacing also matters. If the time interval $\tau$ between the first and last epoch is wide enough it allows reaching the maximum $S_P/S_I$ value. \autoref{f:p_satur} presents the corresponding jet $P$-filling factor. It increases with $\tau$ up to a certain level, which is distributed in a wide range, with a typical value of about 1/3 and then becomes quasi-constant or slightly decreases due to a source evolution. We selected 53 sources with $\tau>10$~yr and $N_\mathrm{epoch}>20$ and found that the time interval, at which the maximum $S_P/S_I$ is reached, ranges from 4 to 17~yr, with a median of $10.5\pm2.7$~yr.

\begin{figure}
    \centering
    \includegraphics[width=\linewidth]{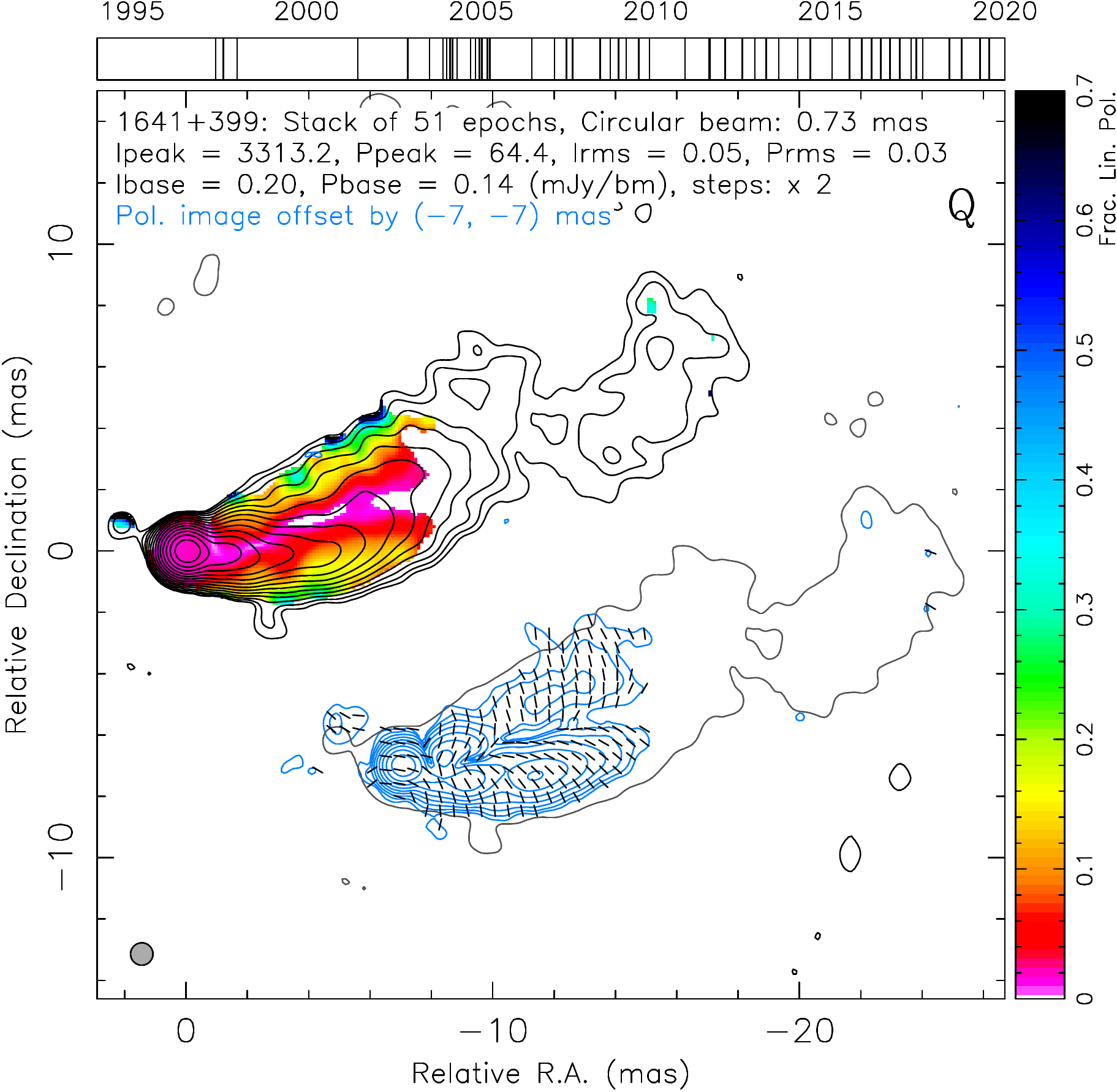}
    \caption{Evolution showing the build-up of the stacked image of the quasar \href{https://www.cv.nrao.edu/MOJAVE/cumulative_stacking_maps/1641+399_stacked_map_evolution.gif}{3C~345} since the epoch of 1997 May 26, continuously adding 50 later epochs until 2019 July 19. Last frame is shown here. Animated version of this figure is available online as supplementary material. 
    \label{f:3c345_gif_ani}
    }
\end{figure}

To visualise how a stacking map evolves by progressively adding more epochs, we made gif-animated cumulative images. In \autoref{f:3c345_gif_ani}, we show the cumulative stacked map of the bright LSP quasar 1641+399 having 51 epochs distributed over 21~yr, as an example. The cumulative stacked maps for other sources are available from the MOJAVE website\footnote{\url{https://www.cv.nrao.edu/MOJAVE/allsources.html}}; select a source of interest and click on 'Cumulative Polarization Stacking' link.

\begin{figure*}
    \centering    
    \includegraphics[width=0.385\linewidth]{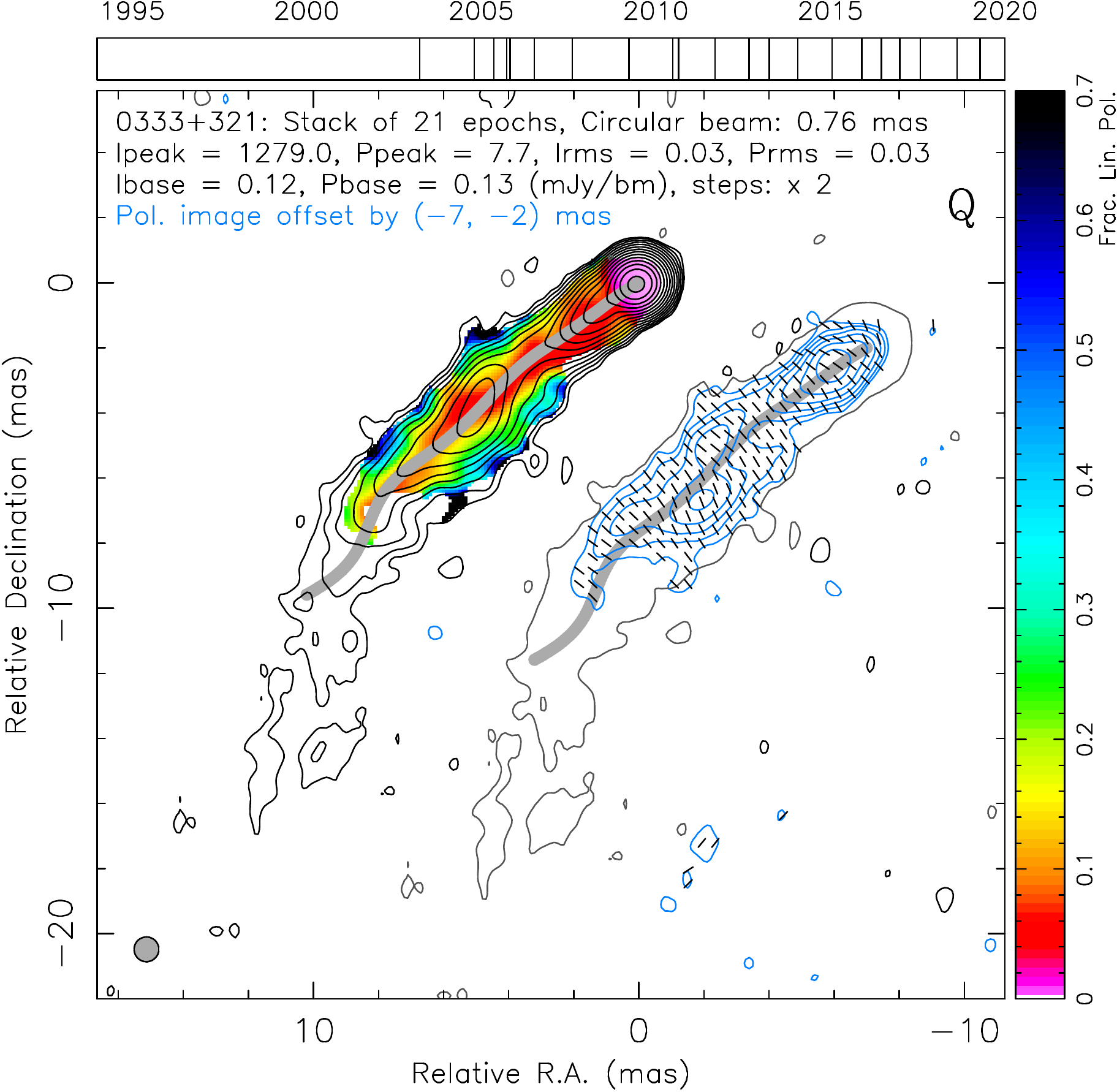}\hspace{0.5cm}
    \includegraphics[width=0.385\linewidth]{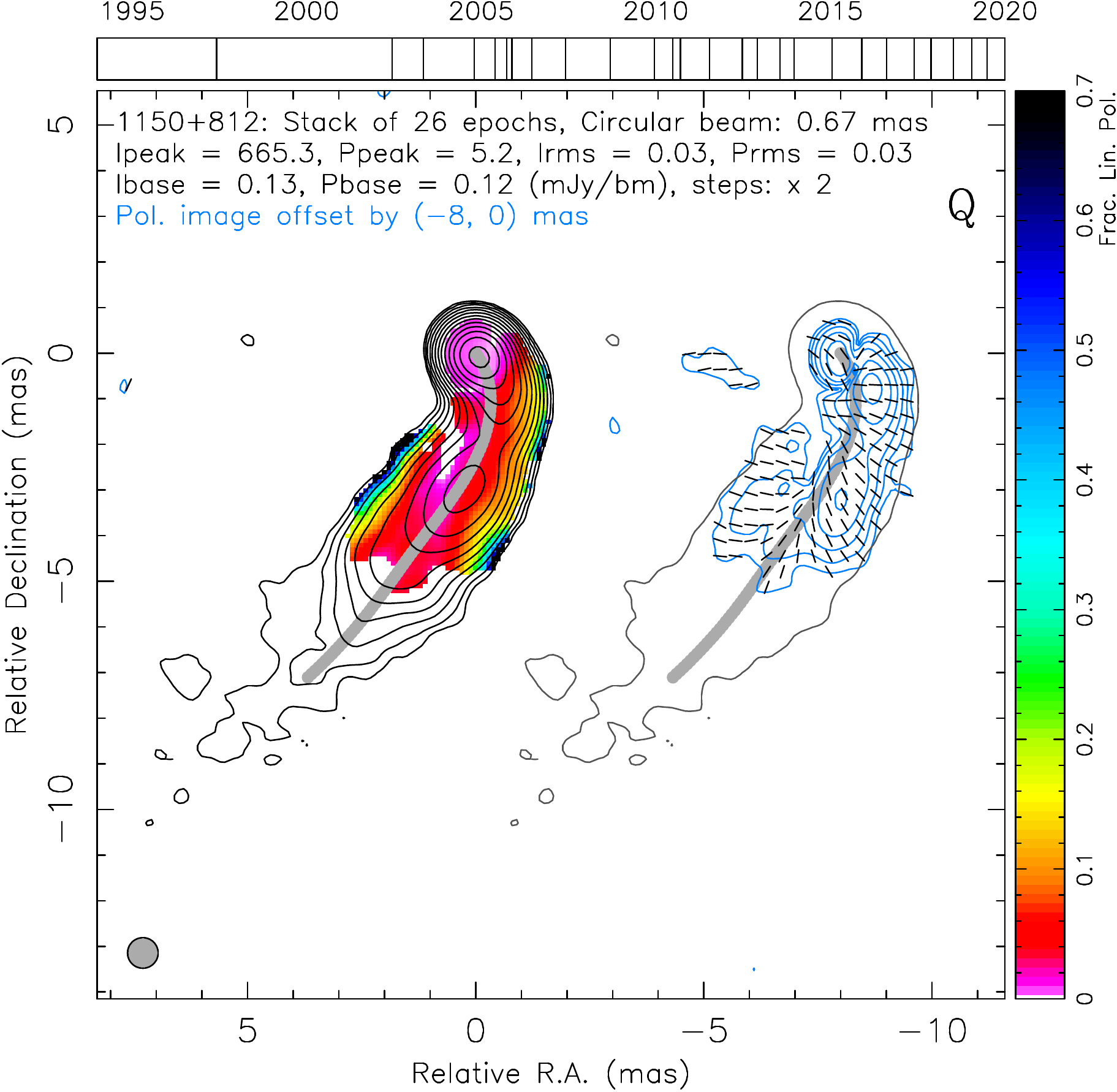}\vspace{0.1cm}
    \includegraphics[width=0.385\linewidth]{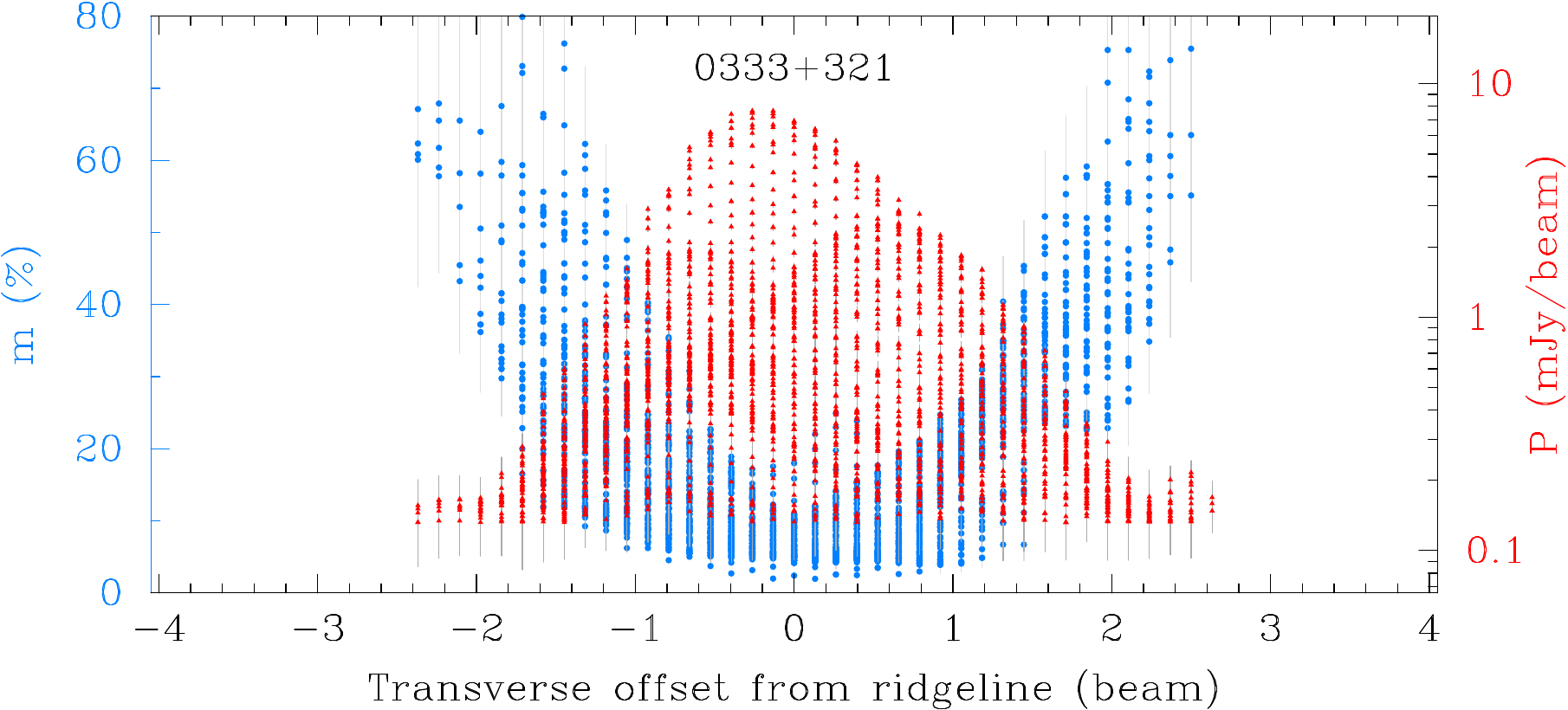}\hspace{0.5cm}
    \includegraphics[width=0.385\linewidth]{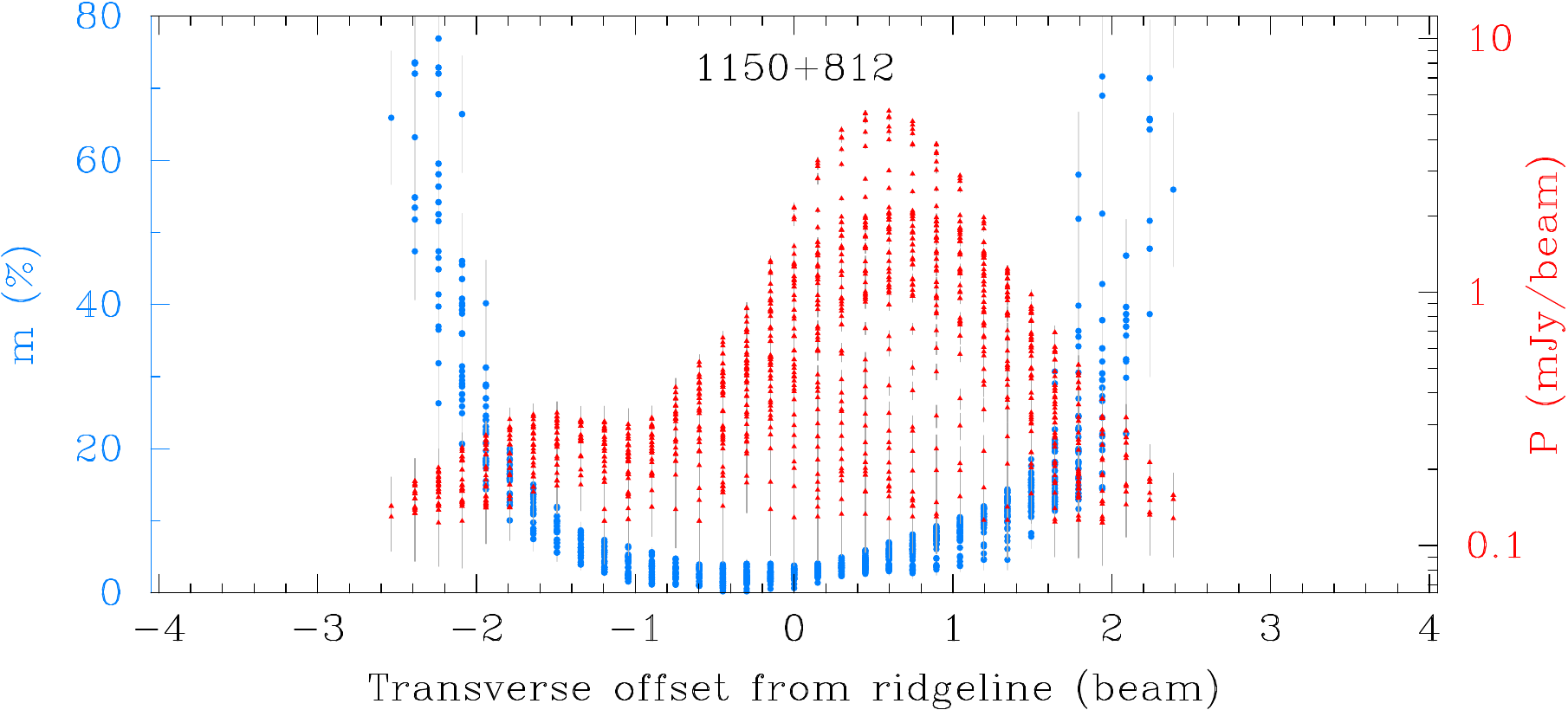}
    \caption{Stacked maps of the quasars 0333+321 and 1150+812 (top panels) showing the U-shaped transverse $m$-profiles (bottom panels; blue dots) and characterized by orthogonal and fountain-like EVPA patterns, respectively. $P$-slices are shown by red triangles. The quasar 1150+812, as well as other sources with the fountain-like EVPA distribution, reveals asymmetric two-peaked $P$-cuts. The profiles are taken excluding the one-beam core region.
    \label{f:U_shaped_m_cut}
    }
\end{figure*}

\begin{figure*}
    \centering    
    \includegraphics[width=0.385\linewidth]{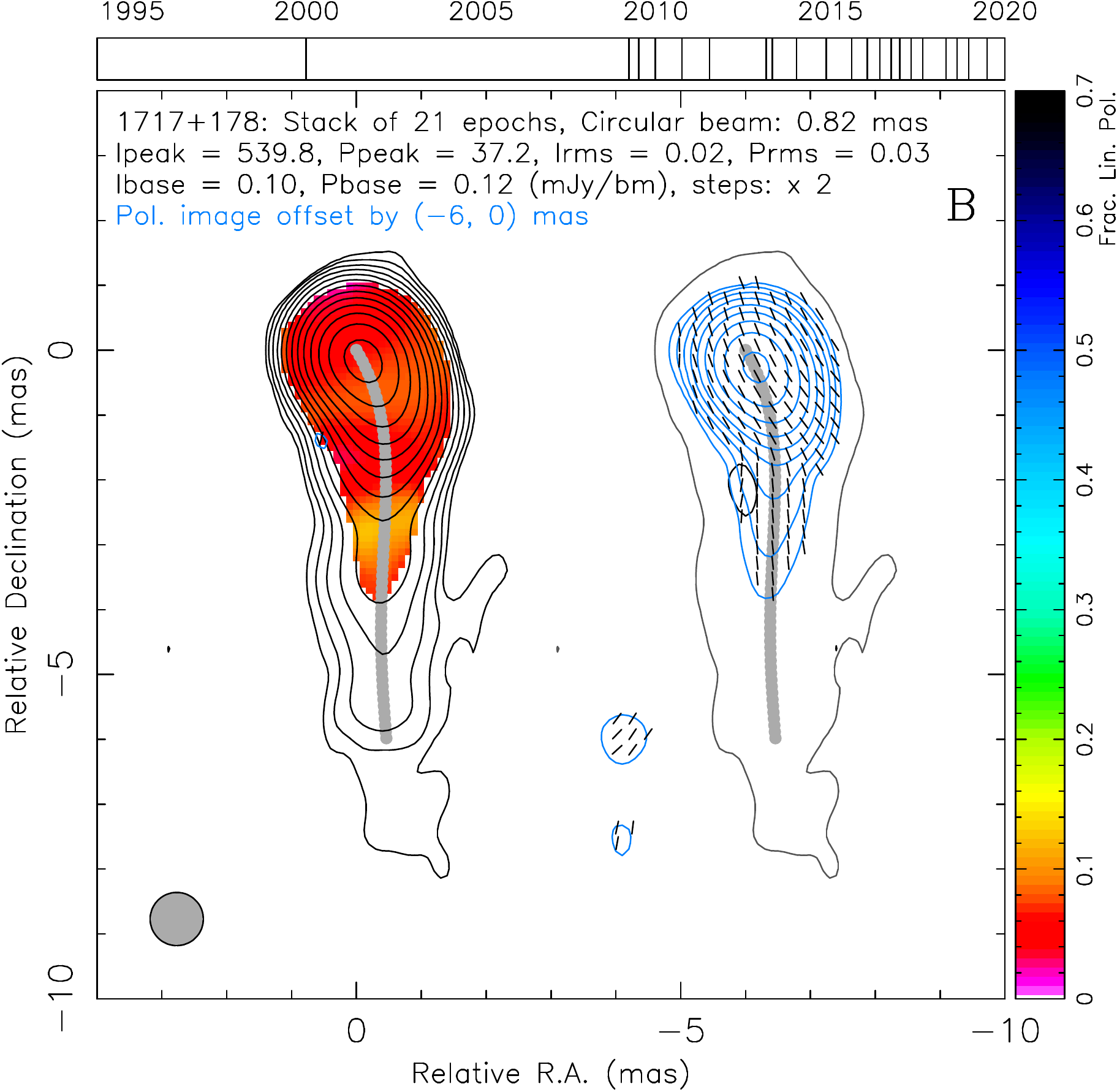}\hspace{0.5cm}
    \includegraphics[width=0.385\linewidth]{figures/fig02_2200+420_stack.pdf}\vspace{0.1cm}
    \includegraphics[width=0.385\linewidth]{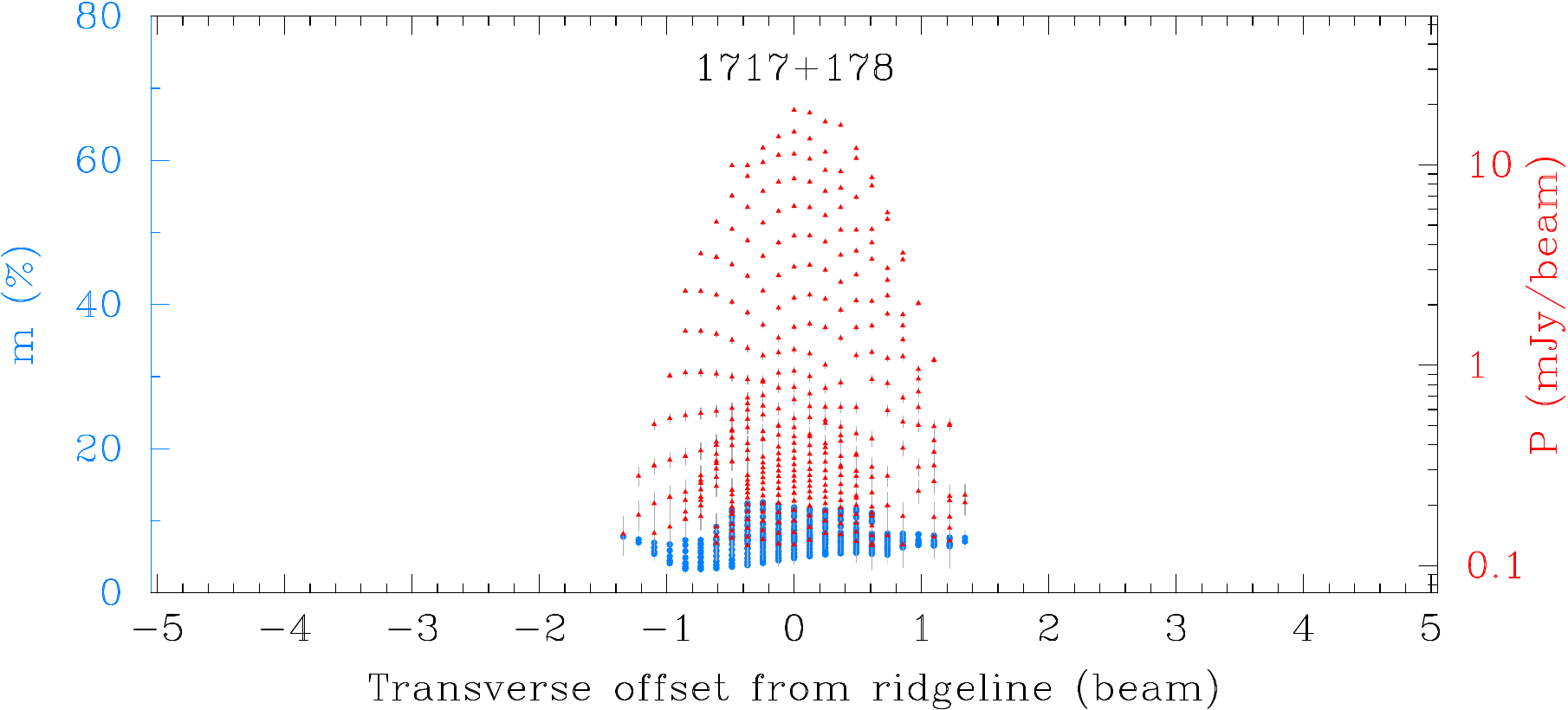}\hspace{0.5cm}
    \includegraphics[width=0.385\linewidth]{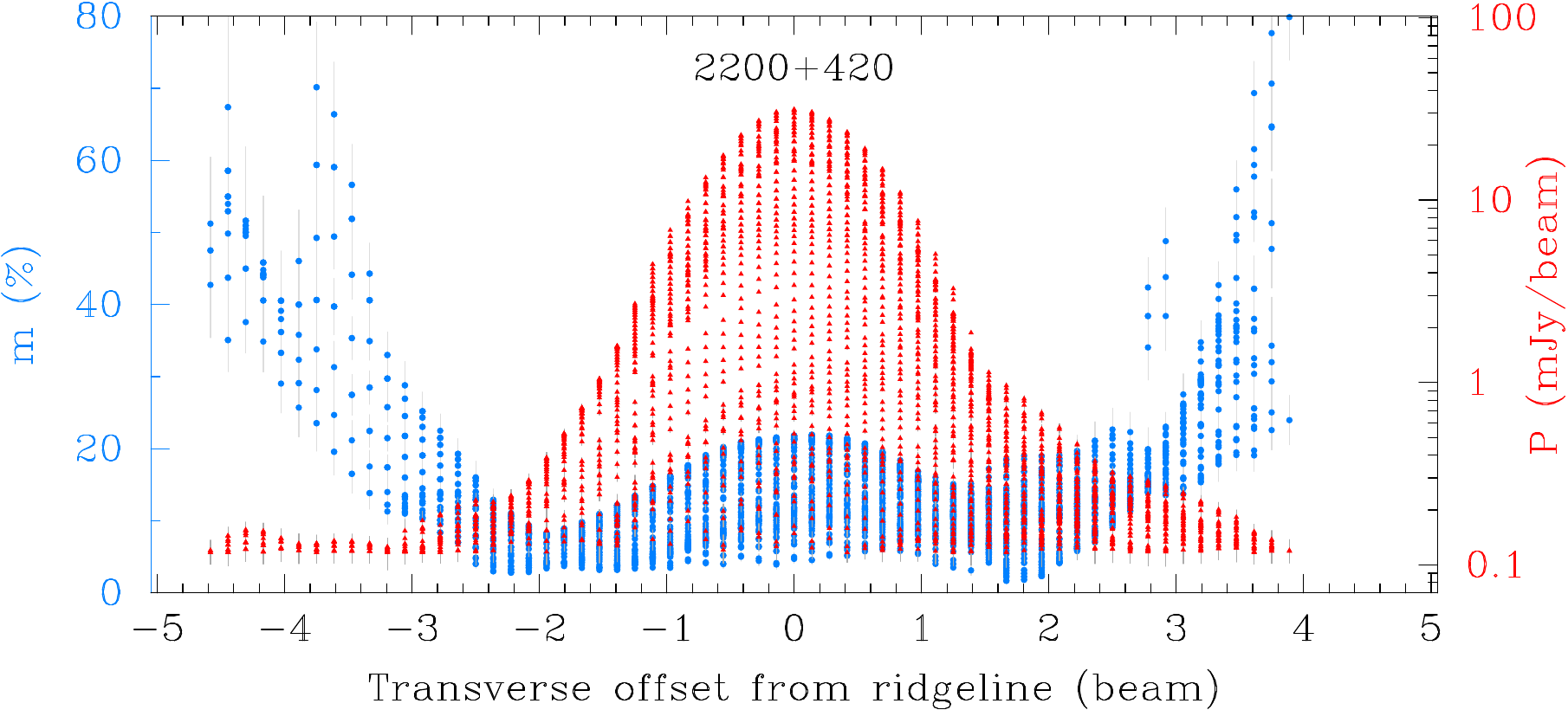}
    \caption{Stacked maps of the BL Lacertae objects 1717+178 and 2200+420 (top panels) showing a quasi-constant and W-shaped transverse $m$-profiles (bottom panels; blue dots), respectively, and characterized by one-peaked $P$-cuts (red triangles). EVPA aligned with local jet direction. The profiles are taken excluding the one-beam core region.
    \label{f:W_shaped_m_cut}
    }
\end{figure*}

\subsection{EVPA patterns and transverse $m$-cut types} \label{s:EVPA_patterns}

\begin{table*}
\centering
\caption{Typical features of the main EVPA patterns and transverse $m$, $P$-cut types revealed in the stacked maps.
\label{t:evpa_patterns_vs_m_cuts}}
\begin{threeparttable}
\begin{tabular}{lcllcc}
\hline\hline\noalign{\smallskip}
$m$-cut type & EVPA pattern & $P$-cut & $P$-signal location & Mainly observed in$^a$ & Examples \\
         (1) &          (2) &     (3) &                 (4) &                    (5) &      (6) \\
\hline
\multirow{2}{5em}{U-shaped}  & \rotatebox[origin=c]{90}{$\prec$}$^b$                 & two-peaked, asymmetric & across entire jet width        &  quasars  &  \href{https://www.cv.nrao.edu/MOJAVE/cumulative_stacking_maps/1641+399_stacked_map_evolution.gif}{1641+399}, \href{https://www.cv.nrao.edu/MOJAVE/cumulative_stacking_maps/1920-211_stacked_map_evolution.gif}{1920$-$211} \\   
                             & $\perp$                                             & one-peaked, asymmetric & across entire jet width        &  quasars  &  \href{https://www.cv.nrao.edu/MOJAVE/cumulative_stacking_maps/1226+023_stacked_map_evolution.gif}{1226+023}, \href{https://www.cv.nrao.edu/MOJAVE/cumulative_stacking_maps/0836+710_stacked_map_evolution.gif}{0836+710} \\ 
                             \cline{2-6}
Flat                         & $\parallel$                                         & one-peaked, symmetric  & close to ridge, never at edges &  BL Lacs  &  \href{https://www.cv.nrao.edu/MOJAVE/cumulative_stacking_maps/1538+149_stacked_map_evolution.gif}{1538+149}, \href{https://www.cv.nrao.edu/MOJAVE/cumulative_stacking_maps/2157+213_stacked_map_evolution.gif}{2157+213} \\ 
                             \cline{2-6}
W-shaped                     & $\parallel$ or \rotatebox[origin=c]{90}{$\prec$}$^b$  & $^c$                   & $^c$                           & BL Lacs / quasars  &  \href{https://www.cv.nrao.edu/MOJAVE/cumulative_stacking_maps/2200+420_stacked_map_evolution.gif}{2200+420}, \href{https://www.cv.nrao.edu/MOJAVE/cumulative_stacking_maps/1611+343_stacked_map_evolution.gif}{1611+343} \\ 
\hline
\end{tabular}
\begin{tablenotes}
\item
$^a$Sources are selected by visual inspection. \\
$^b$Fountain-like EVPA distribution. \\
$^c$Sources with W-shaped $m$-profile might show signatures of U-shaped or flat $m$-cut types.
\end{tablenotes}
\end{threeparttable}
\end{table*}

There are three different EVPA patterns revealed in the stacked maps: (i) with electric vectors transverse to the jet, (ii) aligned with the outflow local direction and (iii) showing a fountain-like distribution, with EVPA following the jet near its axis and gradually rotating to the orthogonal direction at the jet edges. There are three main types of $m$-cuts across the jet, represented by horizontal (quasi-constant $m$), U- or W-shaped profiles. Possible combinations of EVPA patterns and $m$-cut types beyond the core are not equally likely. Thus, the most common case seen in every third source is the U-shaped $m$-slices (\autoref{f:U_shaped_m_cut}), with a dip on the axis and higher values at jet edges. This is accompanied by EVPAs either all transverse to the jet (e.g. 
\href{https://www.cv.nrao.edu/MOJAVE/cumulative_stacking_maps/0214+083_stacked_map_evolution.gif}{0214$+$083},  
\href{https://www.cv.nrao.edu/MOJAVE/cumulative_stacking_maps/0430+052_stacked_map_evolution.gif}{0430$+$052},  
\href{https://www.cv.nrao.edu/MOJAVE/cumulative_stacking_maps/0836+710_stacked_map_evolution.gif}{0836$+$710}) 
or distributed in a fountain-like manner (see, e.g.  
\href{https://www.cv.nrao.edu/MOJAVE/cumulative_stacking_maps/1641+399_stacked_map_evolution.gif}{1641$+$399}, 
\href{https://www.cv.nrao.edu/MOJAVE/cumulative_stacking_maps/1920-211_stacked_map_evolution.gif}{1920$-$211},  
\href{https://www.cv.nrao.edu/MOJAVE/cumulative_stacking_maps/2155-152_stacked_map_evolution.gif}{2155$-$152}).  
This combination of EVPA and $m$-slices is typically seen in quasars with polarized emission detected across the entire width of the jet. Changing of the field order from the centre to the edge of a jet and manifesting U-shaped $m$-cuts can occur naturally in at least two scenarios. First, a turbulent or tangled magnetic field that has been shock-ordered transverse to the flow with shear near the edges of the jet creating a competing longitudinal magnetic field. Second, a helical magnetic field can also produce this pattern as a result of superposition of linearly polarized emission from regions with different EVPA with the most efficient cancellation near the jet axis, depending on the pitch angle (except for special combinations of pitch angle and viewing angle, that in projection can yield a parallel field on one side of the jet and a transverse field on the other \cite[e.g.][]{Lyutikov05}).
Another evidence for the helical B-field scenario is asymmetry of the $m$ and $P$ transverse cuts \citep{Aloy00,Lyutikov05,CB11,Zamaninasab13,Fuentes18,Kramer21}. The latter often show two unequal peaks. We also note that in the case of the fountain-like EVPA pattern, the centre dip in the $m$-profile is deeper than for the cases of purely orthogonal EVPA distribution, sometimes reaching a close to zero level. This kind of EVPA distribution across the jet was obtained by \cite{Murphy13} for a jet model with a helical magnetic field with a relatively large pitch angle $\gtrsim60^\circ$.

\begin{figure}
    \centering
    \includegraphics[width=\linewidth]{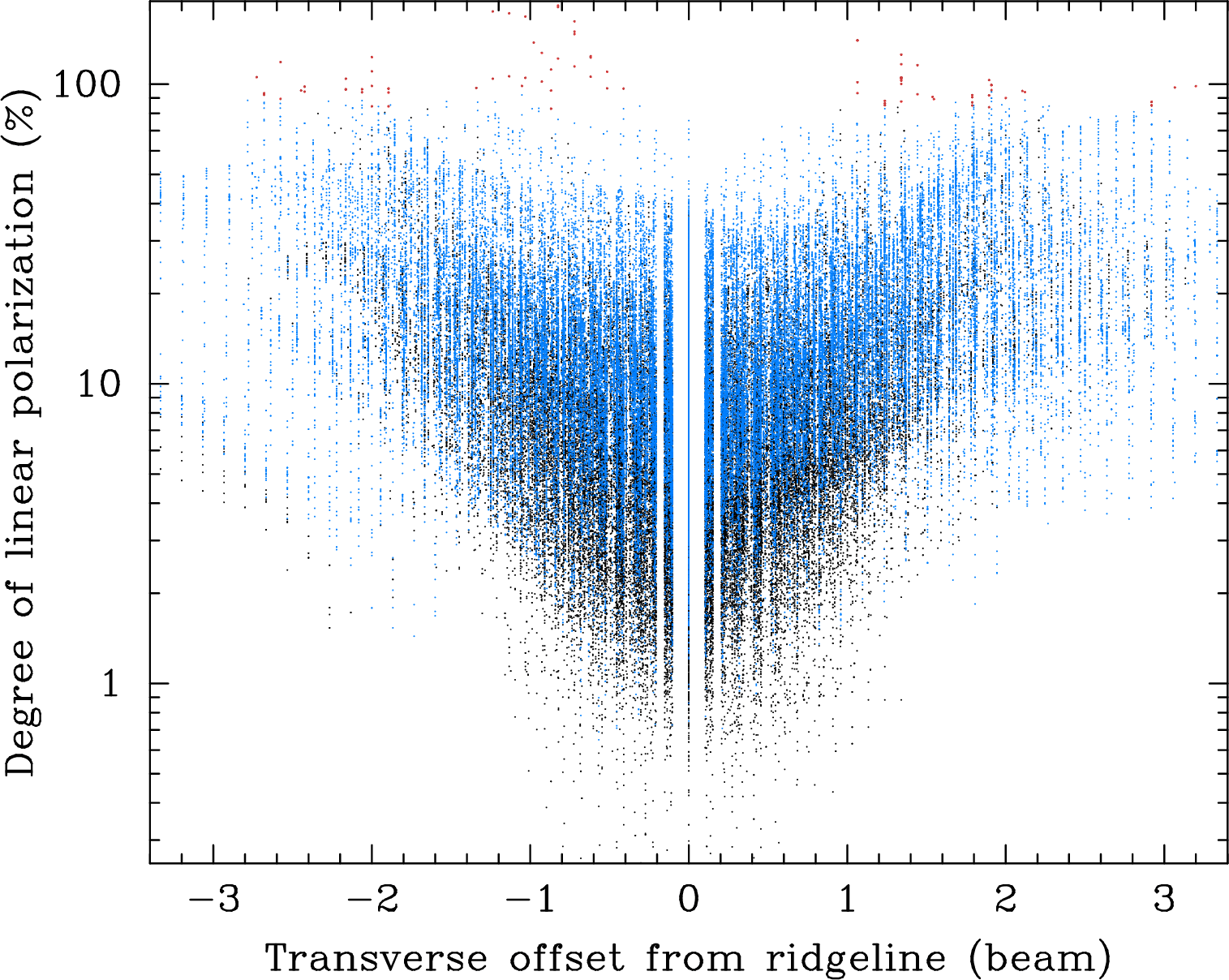}
    \caption{Fractional polarization from slices across the jet as a function of offset (in units of restoring beams) from total intensity ridgeline beyond the one-beam core area for 307 sources, after excluding 58 objects with quasi-constant $m$-profiles. Black dots show measurements from the cuts at distances $r<3$~mas from the core, while light-blue points correspond to those at larger core separations. Red points ($<0.1$ per cent) are outliers, with $m-\sigma_m>75$ per cent.
    \label{f:m_cuts}
    }
\end{figure}

\begin{figure*}
    \centering
    \includegraphics[height=0.32\linewidth]{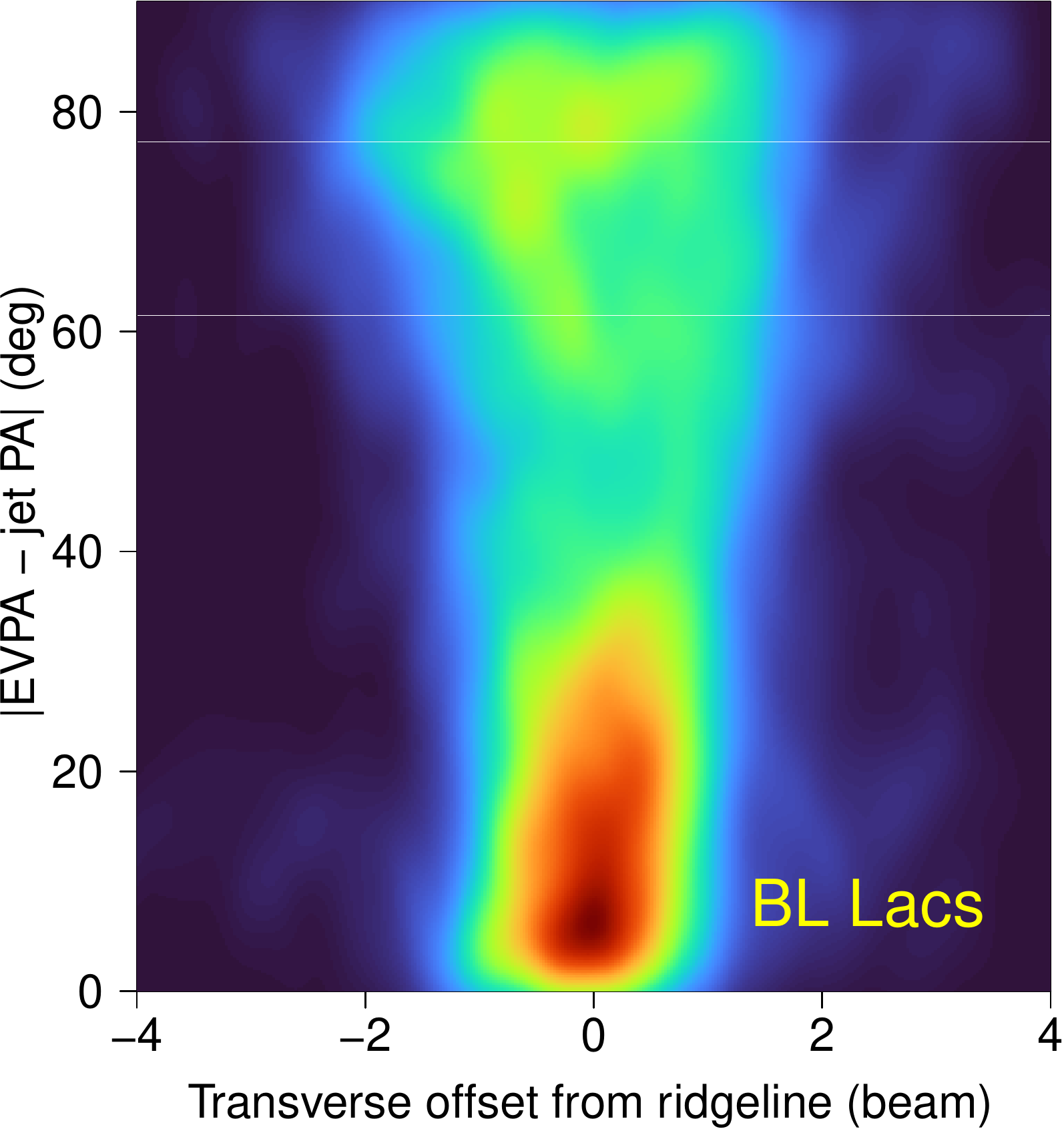}\hspace{0.1cm}
    \includegraphics[height=0.32\linewidth]{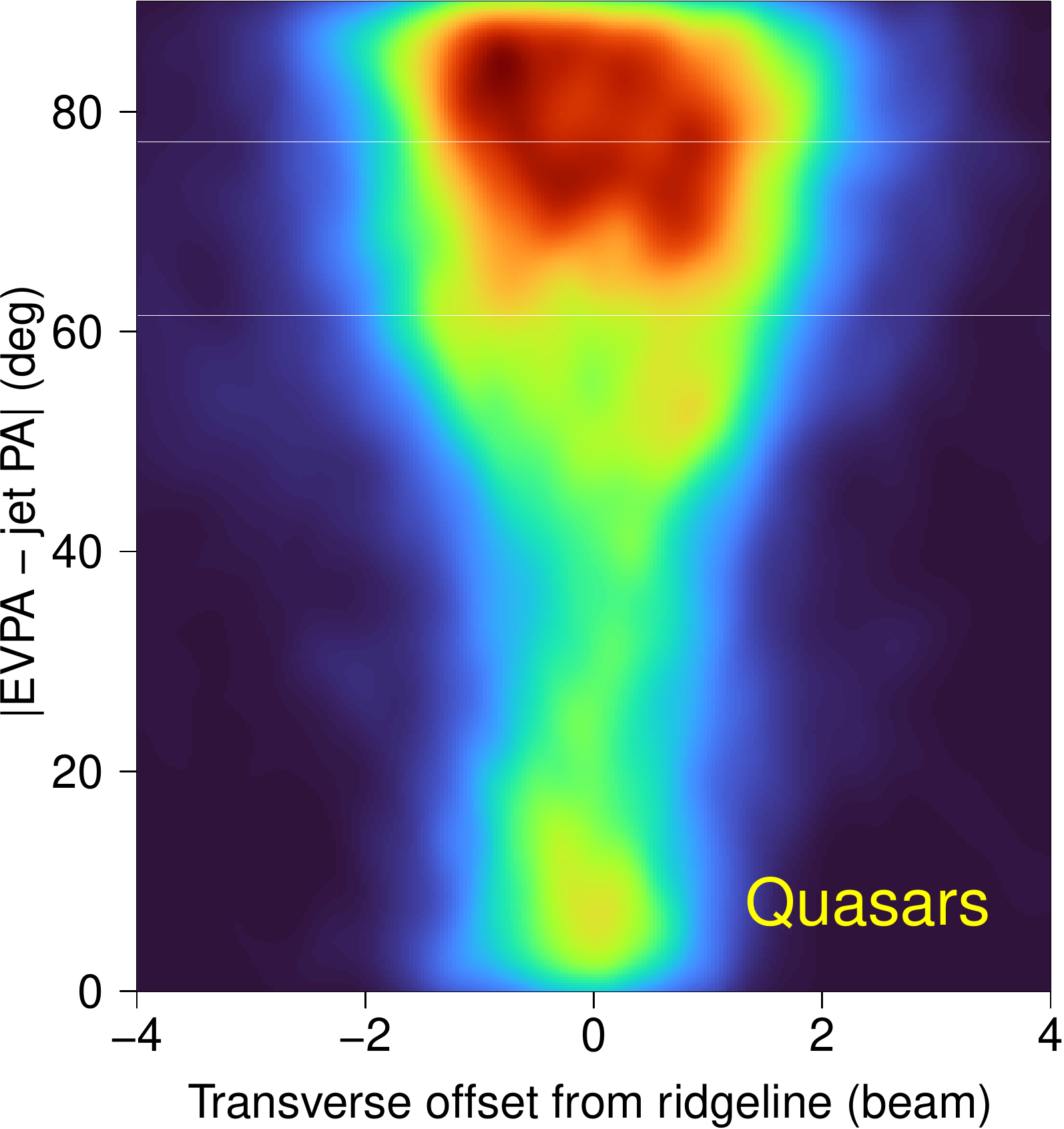}\hspace{0.1cm}
    \includegraphics[height=0.32\linewidth]{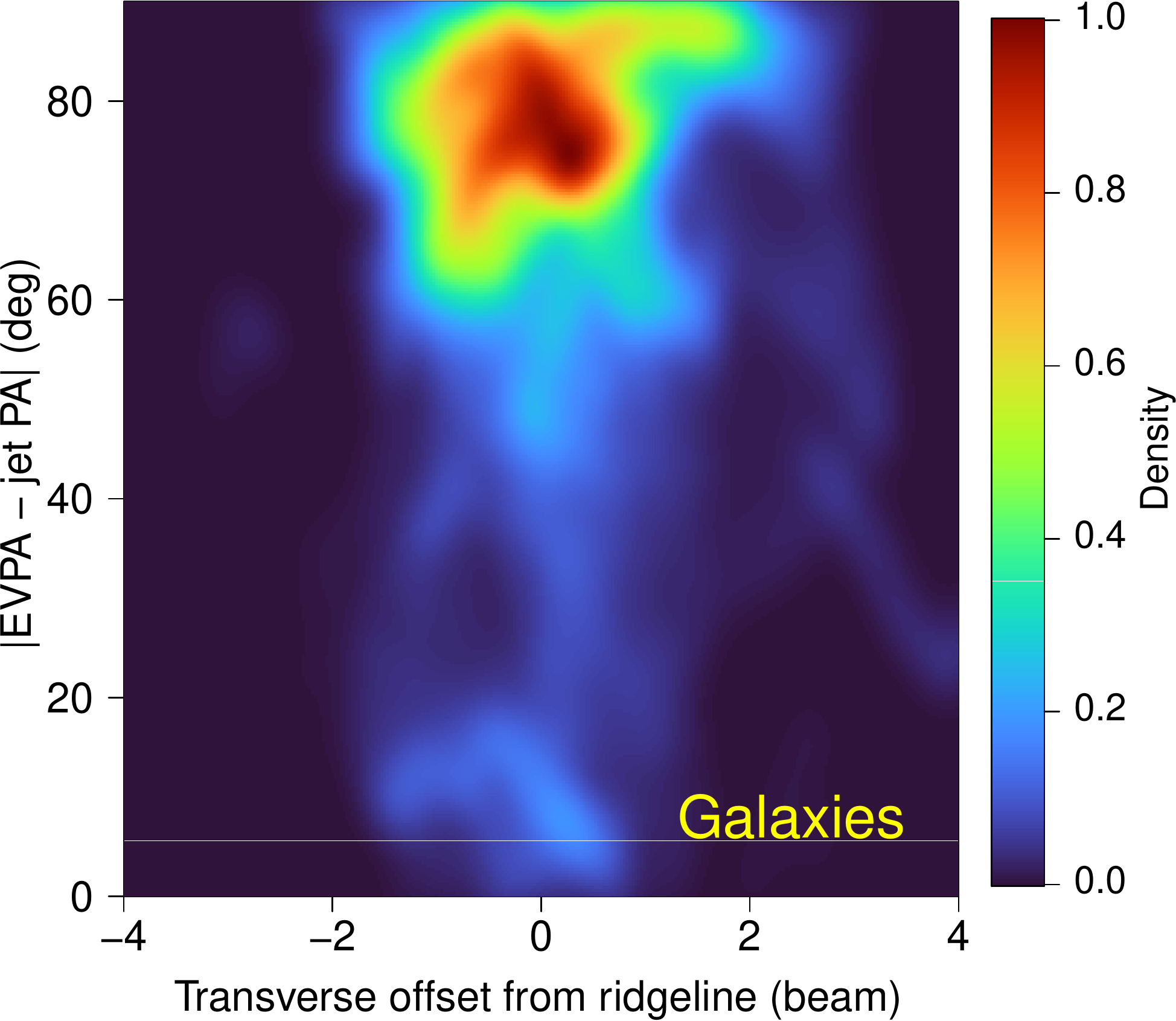}
    \caption{Density distribution of absolute deviation of EVPA from the local jet direction vs transverse offset (in units of restoring beams) from total intensity ridgeline beyond the one-beam core area for 378 sources: 126 BL Lacs (left), 250 quasars (middle) and 12 radio galaxies (right).
    \label{f:evpa_offset_vs_ridge_offset}
    }
\end{figure*}

The second often-seen type of $m$-cut is flat, with no prominent trends (\autoref{f:W_shaped_m_cut}, left). It is accompanied by EVPA well aligned with the jet direction and observed 2.5 times more commonly in BL Lacs (e.g.
\href{https://www.cv.nrao.edu/MOJAVE/cumulative_stacking_maps/0303+490_stacked_map_evolution.gif}{0303$+$490},  
\href{https://www.cv.nrao.edu/MOJAVE/cumulative_stacking_maps/1717+178_stacked_map_evolution.gif}{1717$+$178},  
\href{https://www.cv.nrao.edu/MOJAVE/cumulative_stacking_maps/2157+213_stacked_map_evolution.gif}{2157$+$213}) 
than in quasars (e.g. 
\href{https://www.cv.nrao.edu/MOJAVE/cumulative_stacking_maps/1253-055_stacked_map_evolution.gif}{1253$-$055}, 
\href{https://www.cv.nrao.edu/MOJAVE/cumulative_stacking_maps/1441+252_stacked_map_evolution.gif}{1441$+$252}, 
\href{https://www.cv.nrao.edu/MOJAVE/cumulative_stacking_maps/1754+155_stacked_map_evolution.gif}{1754$+$155}). 
Polarization intensity is quite symmetric and has one peak on the jet axis. Sources with this type of $m$-profile reveal a narrow $P$-distribution in a transverse direction, typically within two beams. The only source showing a flat $m$-profile but EVPA transverse to the jet is the quasar 1700+685.

The third type of $m$-slice is W-shaped (\autoref{f:W_shaped_m_cut}, right). This kind of profile is sort of intermediate between the U-shaped and horizontal $m$-cut types and is quite rare. We visually identified only a few BL Lacs (e.g. 
\href{https://www.cv.nrao.edu/MOJAVE/cumulative_stacking_maps/1307+121_stacked_map_evolution.gif}{1307+121}, 
\href{https://www.cv.nrao.edu/MOJAVE/cumulative_stacking_maps/2200+420_stacked_map_evolution.gif}{2200+420}) 
and quasars (e.g. 
\href{https://www.cv.nrao.edu/MOJAVE/cumulative_stacking_maps/1611+343_stacked_map_evolution.gif}{1611+343}, 
\href{https://www.cv.nrao.edu/MOJAVE/cumulative_stacking_maps/2308+341_stacked_map_evolution.gif}{2308+341}) 
showing this type of $m$-profiles. Their EVPAs follow the jet in the ridgeline area and show progressive rotation to the orthogonal direction at the jet edges if polarization is detected there. Sources with this $m$-cut type show wide jets, up to ten beams in a transverse direction both in total intensity and linear polarization. As shown by \cite{Zakamska08}, this combination of transverse $m$ and EVPA profiles can be produced by a toroidal or highly twisted large-scale B-field.

In \autoref{f:m_cuts}, we plot transverse $m$-cuts beyond the core region for 307 sources except those 58 with quasi-constant $m$-profiles. The cuts are drawn through every ridgeline point typically separated by 0.1~mas. A 10 mas long cut across the jet begins at a point at 5~mas from the ridgeline in $\textrm{PA}_\textrm{cut} = \textrm{PA}_\textrm{jet} - 90^\circ$. The negative offsets are those before the slice crosses the ridgeline. As seen, the dominant fraction of sources from our sample shows a clear increase in fractional polarization towards the jet edges.

We summarise the described connections between the observed EVPA patterns and transverse $m$, $P$-cut types in \autoref{t:evpa_patterns_vs_m_cuts}. In \autoref{f:evpa_offset_vs_ridge_offset}, we present a density distribution of the EVPA orientation with respect to the jet direction for all sources against transverse separation from the ridgeline. It shows that the EVPAs largely either follow the jet within a narrow central channel (typically seen in BL Lacs) or are orthogonal to the outflow across its whole extent, as mainly seen in quasars and radio galaxies.

The observed diversity of polarization patterns listed in \autoref{t:evpa_patterns_vs_m_cuts} can be produced by a helical magnetic field with a different pitch angle and certain geometric and kinematic parameters of the outflow surrounded by a sheath of different thickness and speed relative to the jet spine \citep{BP22}. In particular, one can speculate that in the sources with EVPA aligned with the local jet direction, we detect a narrow stripe of linearly polarized emission around the outflow axis and not at the edges, since it comes from a jet spine, while a sheath is either geometrically thin or its emission is too weak. In contrast, when EVPA is transverse to the jet and polarization is detected across the entire jet width, this can occur when the sheath emission dominates, for example due to its geometric thickness. Indeed, as the jet propagates in the dense environment, the shear layer with a longitudinal B-field (transverse EVPA) can be formed due to interaction with the ambient medium \citep{Laing14}.

Another piece of evidence for the scenario of jet interaction with a dense surrounding medium comes from jet bending. While straight jets typically show quite symmetric $I$ and $P$ cuts, jet bending changes the situation. Out of the visually selected 72 curved jets with detected polarization in the region(s) where the jet is bent, 53 show asymmetry in the transverse $P$-profiles, with an enhanced linearly polarized emission at the outer side of a curved outflow and transverse EVPA in nearly all these cases (see, e.g., \autoref{f:U_shaped_m_cut}, right).

\begin{table}
\centering
\caption{Highly probable neutrino-associated AGN jets observed in the MOJAVE program.
\label{t:neutrino_jets}}
\begin{threeparttable}
\begin{tabular}{lclrrc}
\hline\hline\noalign{\smallskip}
Source & Op. cl. & $z$ & $S_{\rm tot}$ & $P_{\rm tot}$ & Reference \\
       &         &     &         (mJy) &         (mJy) &           \\
   (1) &     (2) & (3) &           (4) &           (5) &       (6) \\
\hline
0506$+$056  &  B  &  0.3365   &   599.1  &  10.3  &  1,2 \\
0735$+$178  &  B  &  \ldots   &   852.7  &  17.2  &  3,4,5,6,7 \\
1253$-$055  &  Q  &  0.536    & 18446.2  & 902.8  &  8 \\
1502$+$106  &  Q  &  1.83786  &  1417.9  &  15.6  &  8 \\
1730$-$130  &  Q  &  0.902    &  3967.0  &  71.0  &  8 \\
1741$-$038  &  Q  &  1.054    &  5803.6  & 109.7  &  8 \\
1749$+$096  &  B  &  0.322    &  3676.9  &  44.7  &  9 \\
2145$+$067  &  Q  &  0.999    &  6371.6  &  91.0  &  8 \\
\hline
\end{tabular}
\begin{tablenotes}
\item
Columns are as follows:
(1) Source name;
(2) Optical class: quasar (Q), BL~Lac (B), radio galaxy (G);
(3) Redshift;
(4) Total cleaned Stokes $I$ flux density in the stacked 15~GHz VLBA image;
(5) Total linear polarization flux density derived from the stacked Stokes $U$ and $Q$ maps as $\sqrt{U^2+Q^2}$;
(6) Reference for neutrino-VLBI association:
1 = \cite{0506_2018}, 
2 = \cite{Baikal0506},
3 = \cite{ATel15098}, 
4 = \cite{ATel15099}, 
5 = \cite{ATel15102}, 
6 = \cite{ATel15105}, 
7 = \cite{ATel15136},
8 = \cite{Plavin20} and
9 = \cite{Plavin22}.
\end{tablenotes}
\end{threeparttable}
\end{table}

\subsection{AGN-neutrino associations}
\label{s:neutrino}

There are eight radio-bright blazars (three BL Lacs and five quasars, see \autoref{t:neutrino_jets}) from our sample, which showed a significant positional association with the high-energy neutrino events detected by IceCube and Baikal-GVD \citep{0506_2018,Plavin20,Plavin22,ATel15098,Baikal0506}. We constructed the most probable list of neutrino associations with VLBI-bright blazars according to the rationale in the references provided. These objects show  different polarization properties and EVPA patterns, from BL Lac 1749+096 with polarization detected only in the central channel to those with a rich polarization distribution across the entire jet, e.g. in 0506+056 or 0735+178. 
Yet, we do not see specific features in linear polarization on parsec scales attributed to the highly probable blazar-neutrino associations.

\section{Summary} \label{sec:summary}

We produced 15~GHz stacked maps in total intensity (I) and linear polarization ($P$, $m$, EVPA) for 436 radio-loud AGN using nearly 6000 full-Stokes, single-epoch VLBA observations, the vast majority of which were carried out within the MOJAVE program. The 368 individual observing epochs cover a time range up to 24 yr, from 1996 January 19 to 2019 August 4. We corrected the $I$ and $P$ stacked maps for a bias induced by the CLEANing procedure. Additionally, the $P$ stacked maps were corrected for the Ricean bias as well. A number of epochs in the stacked maps are distributed from five to 139, with a median of nine. This represents the largest and most complete study of the polarization properties of parsec-scale AGN jets to date.

Our main results are as follows.

1. A stacking procedure based on the Stokes averaging allowed (i) improving image sensitivity by a factor of a few and up to an order of magnitude for the most frequently observed sources and (ii) revealing more complete $I$ and $P$ brightness distributions. We found that stacking over a time span of about ten years is typically enough to delineate the stable and most complete distribution of linear polarization of a source.

2. Fractional polarization shows a significant growth down the jet along its ridgeline traced out in total intensity. This may be caused by a combination of different effects, such as (i) progressively smaller beam depolarization and Faraday rotation with distance to the core, (ii) spectral ageing, (iii) weakening shocks and accompanied turbulence of emitting plasma, (iv) decreasing the pitch angle of a helical magnetic field.

3. The degree of linear polarization measured along the jet axis in BL Lacertae objects is typically higher compared to that of quasars up to hectoparsec de-projected scales. On larger scales, they show comparable values. In particular, LSP BL Lacs and quasars selected from the matched redshift range $z\lesssim0.5$ have no systematic difference in $m$-values along their ridgelines beyond the core. HSP BL Lacs appear distinct from ISP and LSP BL Lacs with lower fractional polarization.

4. The peak of polarization intensity is offset from that of total intensity, with a median shift down the jet of about 0.2~mas for blazars. The shift is mainly caused by opacity in the core, so that the polarization peak position is determined by the innermost jet features, whose emission becomes optically thin. In radio galaxies, the offset is large, with a median of about 3~mas, as the inner jet regions in the galaxies are often subject to strong Faraday depolarization.

5. If polarization is detected over the entire jet width, the fractional polarization increases towards the jet edges, often manifesting as U-shaped profiles across the outflow. In some rare cases, e.g. in BL Lacertae or the quasar 1611+343, W-shaped cuts are detected. Some sources, predominantly BL Lacs, show quasi-constant transverse $m$-cuts.

6. We identified three main EVPA patterns: aligned with ridgeline, transverse and a fountain-like distributions. The latter is an intermediate form between the first two patterns, with EVPA being parallel to the local direction of the jet axis and gradually rotating to the orthogonal direction closer to the jet edges. The EVPA patterns and $m$-cut types are closely connected. Thus, the aligned EVPAs are accompanied by the quasi-constant $m$-profiles seen typically in BL Lacs, while the perpendicular or fountain-like EVPA patterns are tied with the U/W-shaped $m$-slices mostly detected in quasars. The polarization patterns observed in the parsec-scale AGN jets suggest a presence of a helical magnetic field associated with the outflow with a possible shear interaction with the ambient medium. This is further confirmed by a transverse asymmetry of $P$-profiles, with an enhanced $P$-emission typically detected at the outer side of a curved outflow with EVPA transverse to the local jet direction.

The main conclusions listed above also hold for the MOJAVE 1.5 QC flux-density limited sub-sample.

\hfill
\section*{Acknowledgements}
The authors thank the referee John Wardle for constructive comments which helped to improve the manuscript.
We thank members of the MOJAVE team for discussions of the paper, Alan Roy for thorough reading of the manuscript and providing useful comments as well as Elena Bazanova for language editing.
This study was supported by the Russian Science Foundation grant 21-12-00241.
The MOJAVE project was supported by NASA-Fermi grants NNX08AV67G, NNX12A087G and NNX15AU76G.
This work is part of the M2FINDERS project which has received funding from the European Research Council (ERC) under the European Union’s Horizon 2020 Research and Innovation Programme (grant agreement No 101018682).
This research has made use of data from the University of Michigan Radio Astronomy Observatory which has been supported by the University of Michigan and by a series of grants from the National Science Foundation, most recently No.~AST-0607523.
T.S.\ was partly supported by the Academy of Finland projects 274477, 284495, 312496, and 315721.
This research has made use of the NASA/IPAC Extragalactic Database (NED) which is operated by the Jet Propulsion Laboratory, California Institute of Technology, under contract with the National Aeronautics and Space Administration.
The Long Baseline Observatory and the National Radio Astronomy Observatory are facilities of the National Science Foundation operated under cooperative agreement by Associated Universities, Inc. 
This work made use of the Swinburne University of Technology software correlator \citep{DiFX}, developed as part of the Australian Major National Research Facilities Programme and operated under licence.

\section*{Data Availability}
The fully calibrated single-epoch images of Stokes parameters $I$, $Q$ and $U$ at 15~GHz from the MOJAVE program as well as the constructed $I$, $P$, $m$, and $EVPA$ stacked maps in FITS format are available online\footnote{\url{https://www.cv.nrao.edu/MOJAVE}}. The modelfitted structure of AGNs is taken from \citet{MOJAVE_XVIII}.

\bibliographystyle{mnras}
\bibliography{ms_refs}

\appendix

\section{Error estimation: D-terms, I, P, EVPA from simulations and analytical expressions}
\label{a:errors}
The uncertainty estimation of VLBI images is a complicated issue. This is because the procedure for VLBI data processing is complex and includes numerous non-linear transformations and many sources of noise. Also, pixel values are correlated due to the beam convolution. Moreover, \cite{Hovatta_2012} found that the CLEAN procedure itself produces an error that could depend on the source structure. To estimate the uncertainties of the stacked images of the total intensity, linearly polarized flux, EVPA and fractional polarization, we employed both an analytical approach and Monte Carlo (MC) simulations.

\subsection{Monte Carlo simulations}
\label{a:mc_sim}
\citet{2019MNRAS.482.1955P} found that a simulation-based approach is statistically more optimal than the conventional approach of \cite{Hovatta_2012} for single-epoch images. Moreover, as polarized intensity follows the Rice distribution \citep{Vinokur65,2017isra.book.....T}, obtaining analytical estimates of errors for average images of polarized intensity is problematic. The main idea of the method is to replicate the observed data sets a large number of times using source and noise models and assess the impact of different noise realizations across data replications on the resulting stacked map. Here, we follow \citet{2019MNRAS.482.1955P} but with multi-epoch data sets and a more sophisticated noise model. 

\begin{figure}
\centering
\includegraphics[width=\columnwidth]{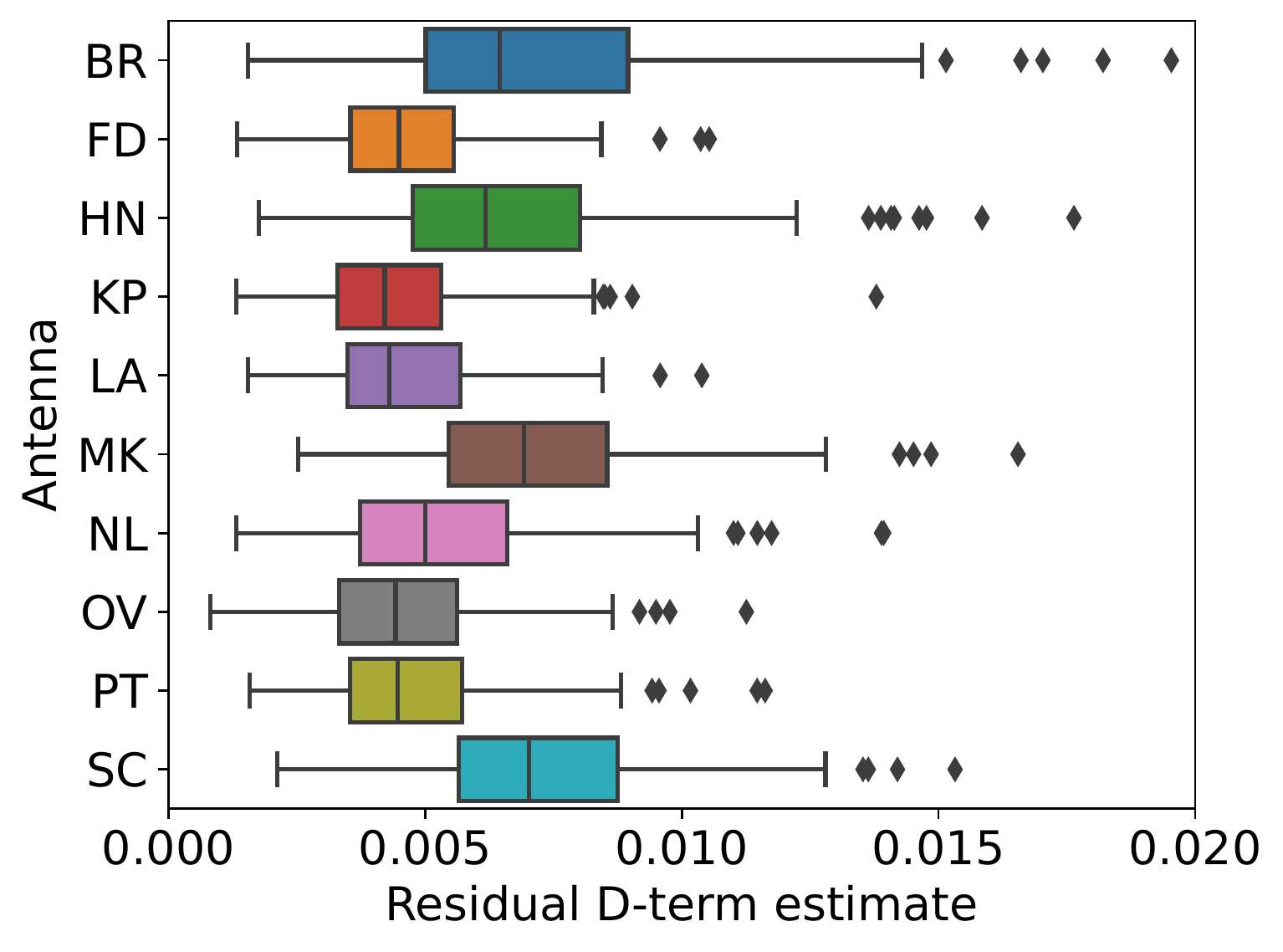}
\caption{Distribution of the residual D-term amplitude for each VLBA station. The residual D-term amplitude was estimated for each antenna from the standard deviations of the difference between the individual source solutions and their median value for the real and imaginary part of the D-term for each IF and polarization. The boxes extend from the first quartile to the third quartile of the data, with a line at the median. The whiskers extend from the box by 1.5x the inter-quartile range. The outlier points are those past the end of the whiskers. Twelve points with values up to 0.063 are not shown.}
\label{fig:resdterms}
\end{figure}

Our noise model accounts for the thermal noise, the residual amplitude scale uncertainty, the residual instrumental (D-terms) uncertainty and the uncertainty of the absolute EVPA of linear polarization. We assume the thermal noise to be Gaussian distributed in the real and imaginary parts of the visibilities. Its value at each baseline was estimated from the observed visibilities with a successive differences approach \citep{briggs}. The residual amplitude scale uncertainty arises from the uncertainty of the antenna gains amplitudes left after self-calibration. It was estimated as 5~per~cent by \citet{Hovatta_2014}. This corresponds to the uncertainty of the gains amplitude for the left and right polarization at a given antenna $\sigma_{g,{\rm R,L}} = 0.035$. We estimated the residual D-term value $\sigma_{{\rm ant},i}$ from the scatter of the D-term solutions for each source in the MOJAVE experiments for each $i$-th antenna. The distribution of the amplitudes of the residual D-terms is presented for all VLBA antennas in \autoref{fig:resdterms}. For non-VLBA antennas (VLA single station at eight epochs), we used the median value for the VLBA antennas. The uncertainty of the absolute EVPA of the linear polarization $\sigma_{\rm EVPA} = 3^\circ$ was estimated in \citep{Hovatta_2012}.

To obtain the artificial visibility data set at a current epoch in a single MC realization, we do the following steps.

\begin{enumerate}
    \item Create the model visibilities, using I, Q and U CLEAN models and $uv$-coverage obtained from the real data set. Here, we shifted the models to put the core position at the phase centre estimated by fitting a model of Gaussian components to visibilities \citep{MOJAVE_XVIII}.
    \item Add a thermal noise.
    \item Scale amplitudes of the parallel and cross-hand correlation by a random factor, corresponding to a 5~per~cent amplitude scale uncertainty. For this, draw two random scale factors $C_R$ and $C_L$ (corresponding to $R$ and $L$ antenna polarization) from the Normal distribution $N(0, 0.035)$ and multiplied the cross-hand product $XY$ by a factor $C_X C_Y$, where $X$ and $Y$ denotes any of ${R, L}$ polarization.
    \item Draw the real and imaginary part of the residual D-term for each polarization ($R$, $L$) for each IF for the $i$-th antenna from a Normal distribution $N(0, \sigma_{\rm D})$, where $\sigma_{\rm D}$ were estimated from the scatter of the D-term solutions, as described above.
    Add the obtained residual D-terms into the data, using the linear approximation \citep{Roberts_1994}.
    \item Rotate the EVPA by a random value drawn from $N(0, \sigma_{\rm EVPA})$. The rotation is done in the $uv$-plane by applying the same value to all Stokes $Q$ and $U$ visibilities.
\end{enumerate}
The obtained artificial data sets were imaged in \texttt{Difmap}, using the same script as the observed data. Then, stacking was performed as with the real observed data \autoref{sec:stacked_maps}. We estimated the dispersion (i.e. the random error) at each single pixel by the scatter of the pixel values across MC realisations. 

\subsection{Analytical uncertainty estimation}
\bigskip
The errors of the linear polarization intensity are estimated as

\begin{equation}
\sigma_P = \frac{\sigma_Q+\sigma_U}{2}\,,
\end{equation}

\noindent
where $\sigma_Q$ and $\sigma_U$ were calculated using the relations from \cite{Hovatta_2012},
taking into account that (i) the Stokes $Q$ and $U$ image rms, instrumental polarization and CLEAN procedure contribute to the error budget, (ii) we deal with the maps stacked over $N_{\rm epochs}$, (iii) typical amplitude of residual D-terms is about 0.005 (\autoref{fig:resdterms}):

\begin{equation}
\sigma_{Q,U} = (\sigma_{\rm rms}^2 + \sigma_{\rm Dterm}^2 + \sigma_{\rm CLEAN}^2)^{1/2}\,,
\end{equation}

\noindent
where

\begin{equation}
\sigma_{\rm Dterm} = 0.005\left(\frac{I^2 + (0.3\times I_{\rm peak})^2}{N_{\rm ant}\times N_{\rm IF}\times N_{\rm scan}\times N_{\rm epoch}}\right)^{1/2}\,,
\end{equation}

\begin{equation}
\sigma_{\rm CLEAN} = 1.5\sigma_{\rm rms}/N_{\rm epoch}
\end{equation}

Thus, the $P$-errors are non-uniformly distributed over a map, with a maximum in the brightest region, which typically represents the VLBI core position. The values $N_{\rm ant}=10$, $N_{\rm IF}=8$ $N_{\rm scan}=4$ were adopted. For $Q$ and $U$ maps, the corresponding $\sigma_{\rm rms}$ was calculated as a median after excluding the highest value of rms from four image corner quadrants. Neglecting the covariance between $I$ and $P$ in a given pixel and taking into account the uncertainty in the antenna gains amplitude at a level of 5~per~cent, the errors in fractional polarization were estimated as

\begin{equation}
\sigma_m = m\left[\left(\frac{\sigma_P}{P}\right)^2 + \left(\frac{\sigma_I}{I}\right)^2 + \frac{0.05^2}{N_{\rm epoch}} \right]^{1/2}\,.
\label{e:m_err_an}
\end{equation}

\begin{figure}
    \centering
    \includegraphics[width=0.95\columnwidth]{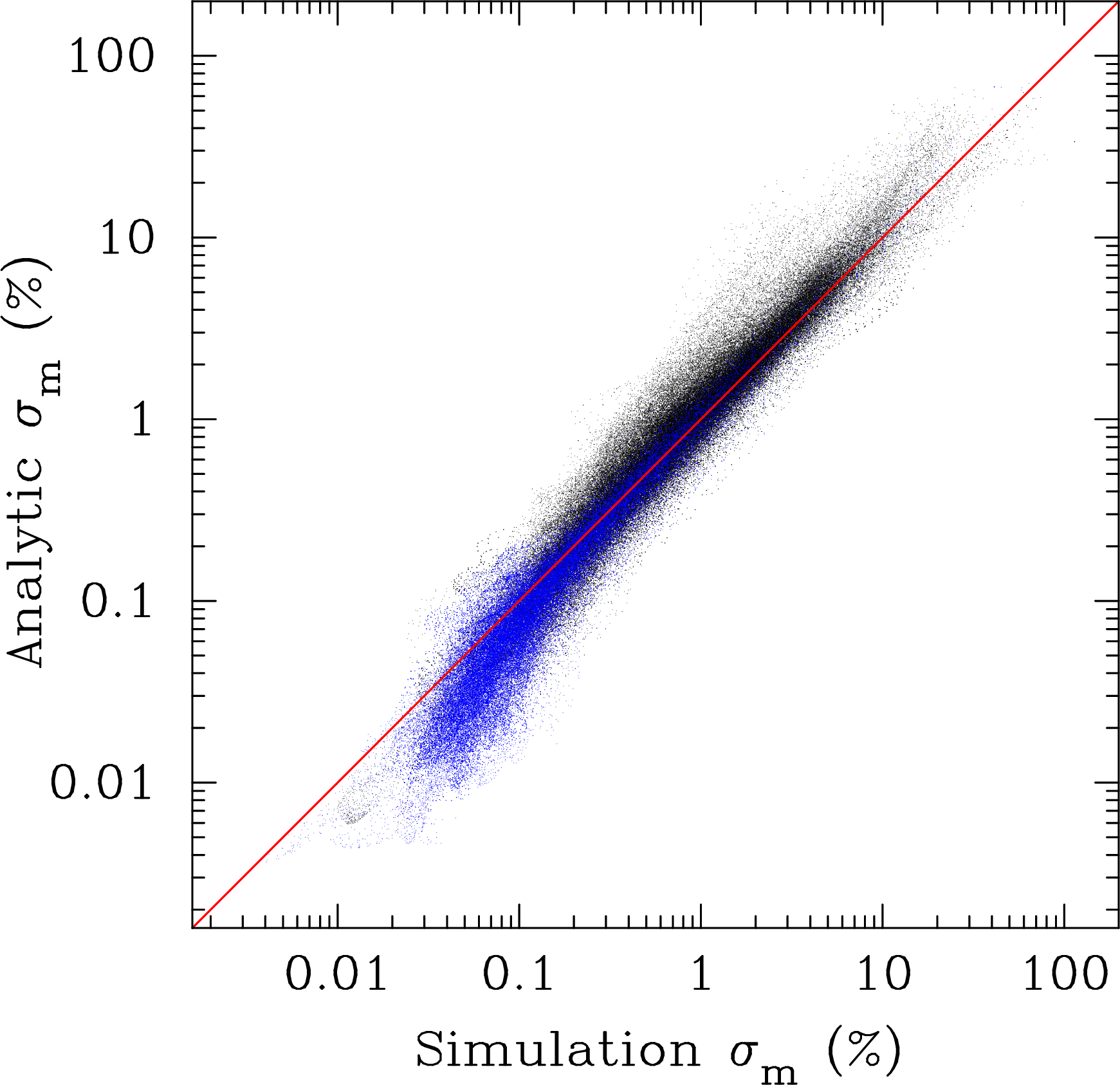}
    \caption{Uncertainties of fractional polarization calculated from the analytic expression 
             (\autoref{e:m_err_an}) against those derived from simulations for 424 sources. 
             The equality line is shown by red. The blue points represent the core area. 
             Good agreement is present in a range over two orders of magnitude, with some 
             underestimation of the analytic $\sigma_m$ for the core regions typically 
             characterised by small values of $m$.
    \label{f:m_err}
    }
\end{figure}

\subsection{Comparing simulation based and analytical uncertainty estimates}
We compared both approaches of the uncertainty estimation using the stacked images of the fractional polarization.
In general, the uncertainty estimates agree very well over a wide range of fractional polarization levels \autoref{f:m_err}. However, the Monte Carlo approach reveals a somewhat larger error near the minimum observed fractional polarization levels. In the regions with low fractional polarization, i.e. the core regions with a high signal-to-noise ratio (SNR), the errors are dominated by the residual instrumental polarization and the residual uncertainty in the amplitude scale of the antenna gains. In the conventional approach, taking these factors into account is problematic.

\section{CLEANing bias}
\label{a:clean_bias}

Stacking increases the sensitivity, i.e. decreases the random error. This raises question about possible systematics or bias that could affect stacked images. From our simulations we also estimated the bias (i.e. systematic error) as a difference between the mean of the images obtained from MC realizations and `true' value.  To create the `Ground Truth' images, we employed CLEAN models which were used to create MC realisations (see item (1) in Appendix~\ref{a:errors}), convolved with a CLEAN beam. Bias estimation is not feasible within the conventional approach \citep{Hovatta_2012} as it requires knowledge of the true model.

\begin{figure*}
    \centering
    \includegraphics[width=0.49\linewidth]{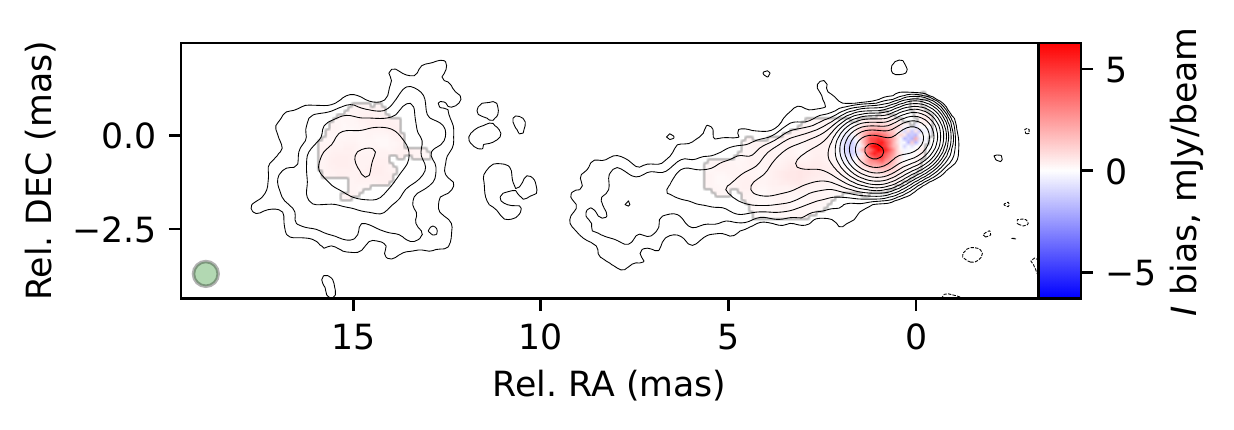}\hspace{0.1cm}
    \includegraphics[width=0.49\linewidth]{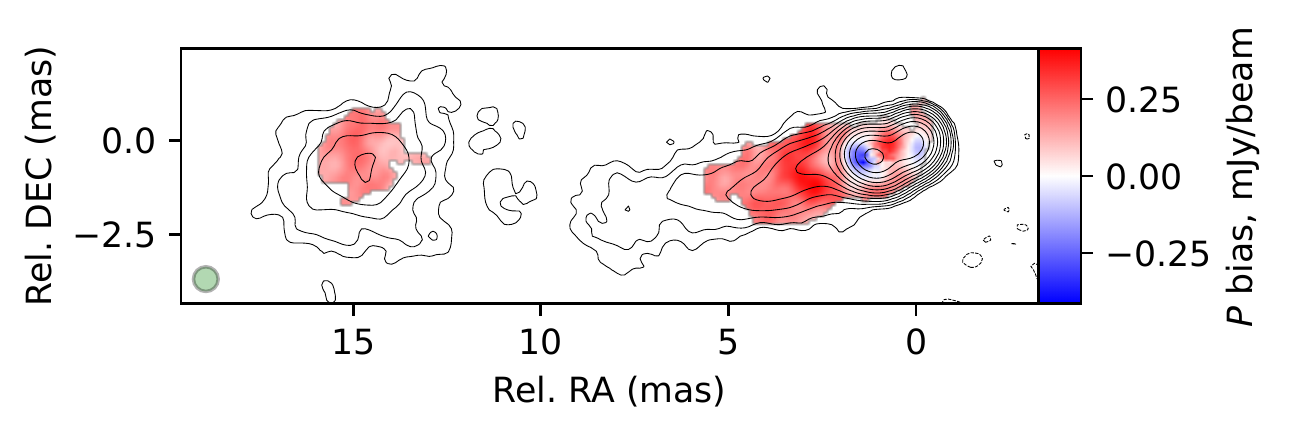}\vspace{0.01cm}
    \includegraphics[width=0.49\linewidth]{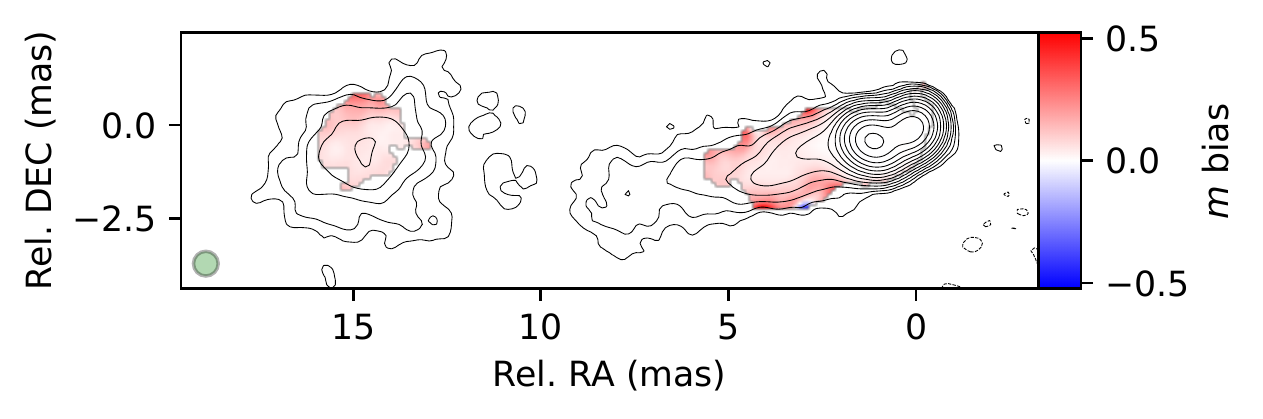}
    \includegraphics[width=0.49\linewidth]{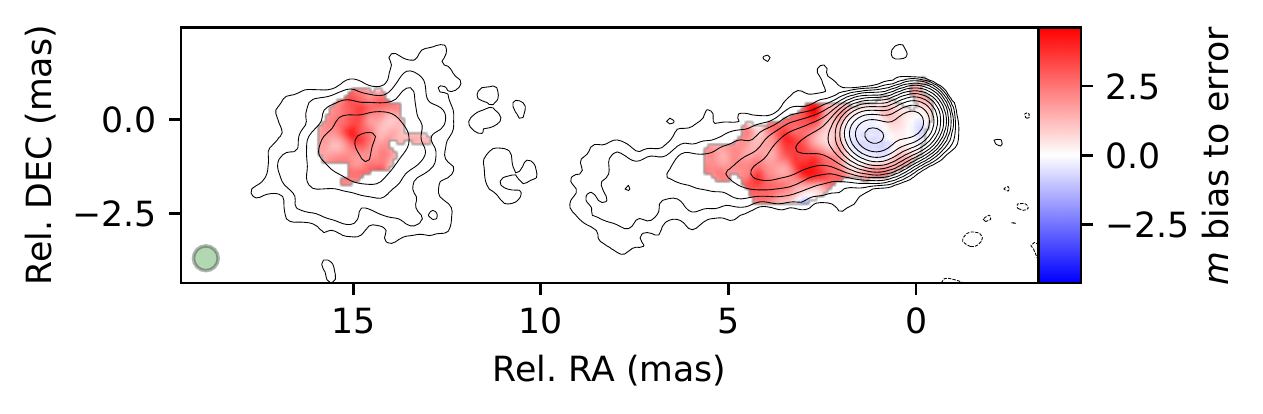}
    \caption{Bias images for source 0212$+$735 for total intensity (top left), linear polarized intensity (top right) and fractional polarization (bottom left) stacked maps. Also, the ratio of the bias to random error is presented for fractional polarization (bottom right).}
    \label{f:bias_images}
\end{figure*}

The images of the bias for total intensity, linearly polarized intensity and fractional polarization are shown in \autoref{f:bias_images} for the source 0212$+$735 as a typical example, having 11 observing epochs. We found that the CLEAN images of EVPA are practically unbiased, most likely due to symmetry, as there is no reason for EVPAs to be biased systematically in one direction or another. The most striking is the increase of the fractional polarization at the jet edges. Previous works on the increase of $m$ towards the jet edges: Mrk\,501 \citep{2005MNRAS.356..859P,Murphy13} and 3C\,120 \citep{2008ApJ...681L..69G} interpreted this as a manifestation of a helical magnetic field. This is because $m$ depends on the component of the magnetic field in the plane of the sky, which is enhanced at the jet edges for a helical field \citep{2014MNRAS.444..172G}. A trend of increasing $m$ with distance in the jets has been also detected by many studies; see details in \autoref{s:m_vs_r}.

The bias in $m$ is determined by the bias of the linearly polarized flux. Experimenting with various CLEAN parameters, we found that deep CLEAN significantly reduces the biases. This suggests that the origin of the bias is the residual (i.e. uncleaned) flux. Indeed, a CLEAN image contains CLEAN components convolved with a CLEAN beam and the residuals effectively convolved with a dirty beam. Size mismatch between the CLEAN and dirty beam changes the relative flux scales of the CLEANed and residual flux. The solution is to CLEAN deep into the noise \citep{briggs,1999ASPC..180..301F}. We tested it by CLEANing deeper by a factor of three and obtained reduced bias levels. This is consistent with results of \cite{2023arXiv230112861P}, who studied various types of systematics in the CLEAN spectral maps and found that shallow CLEAN generally steepens the spectrum in the extended low brightness jet regions.

\begin{figure*}
    \centering
    \includegraphics[width=0.481\linewidth]{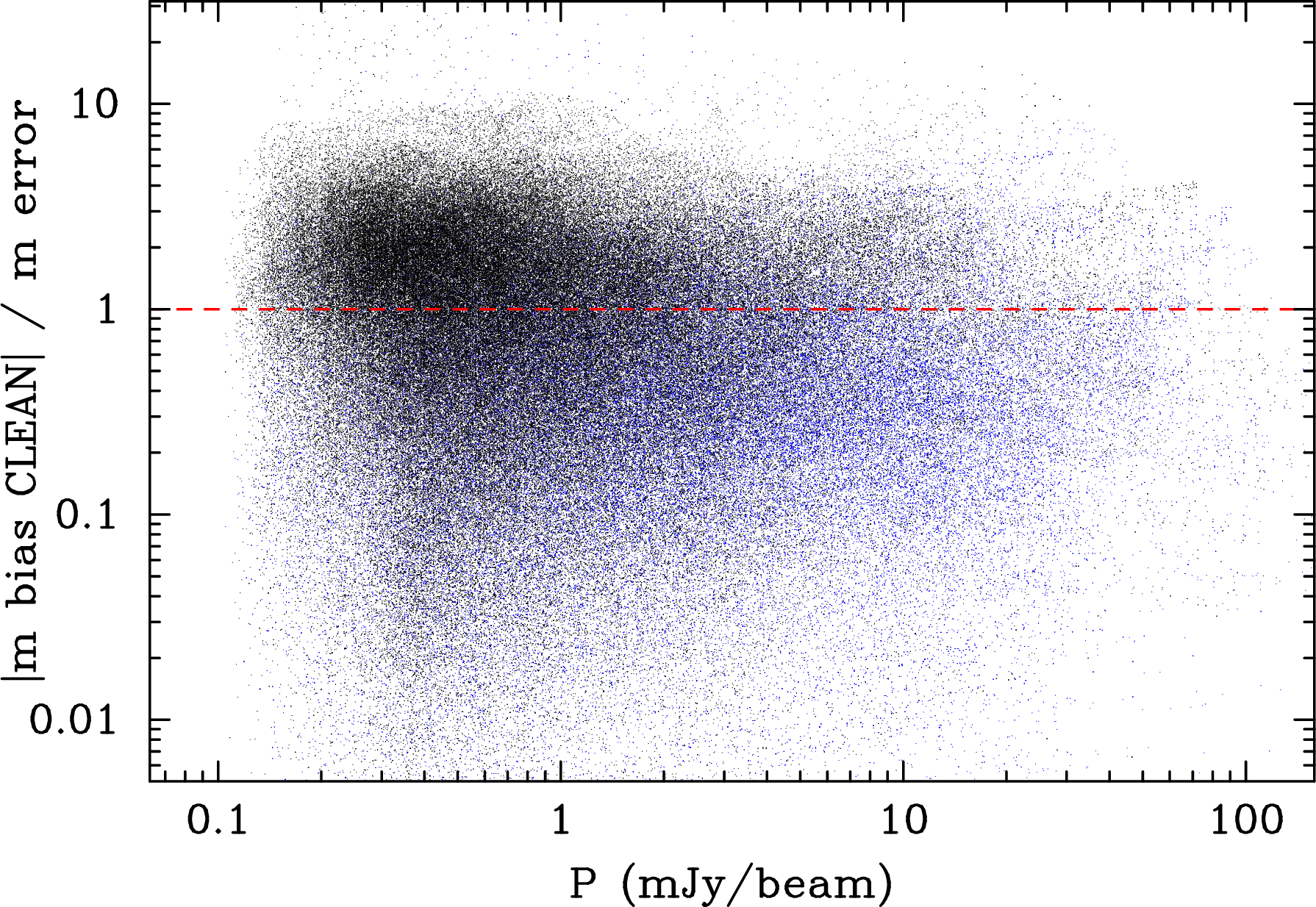}\hspace{0.8cm}
    \includegraphics[width=0.35\linewidth]{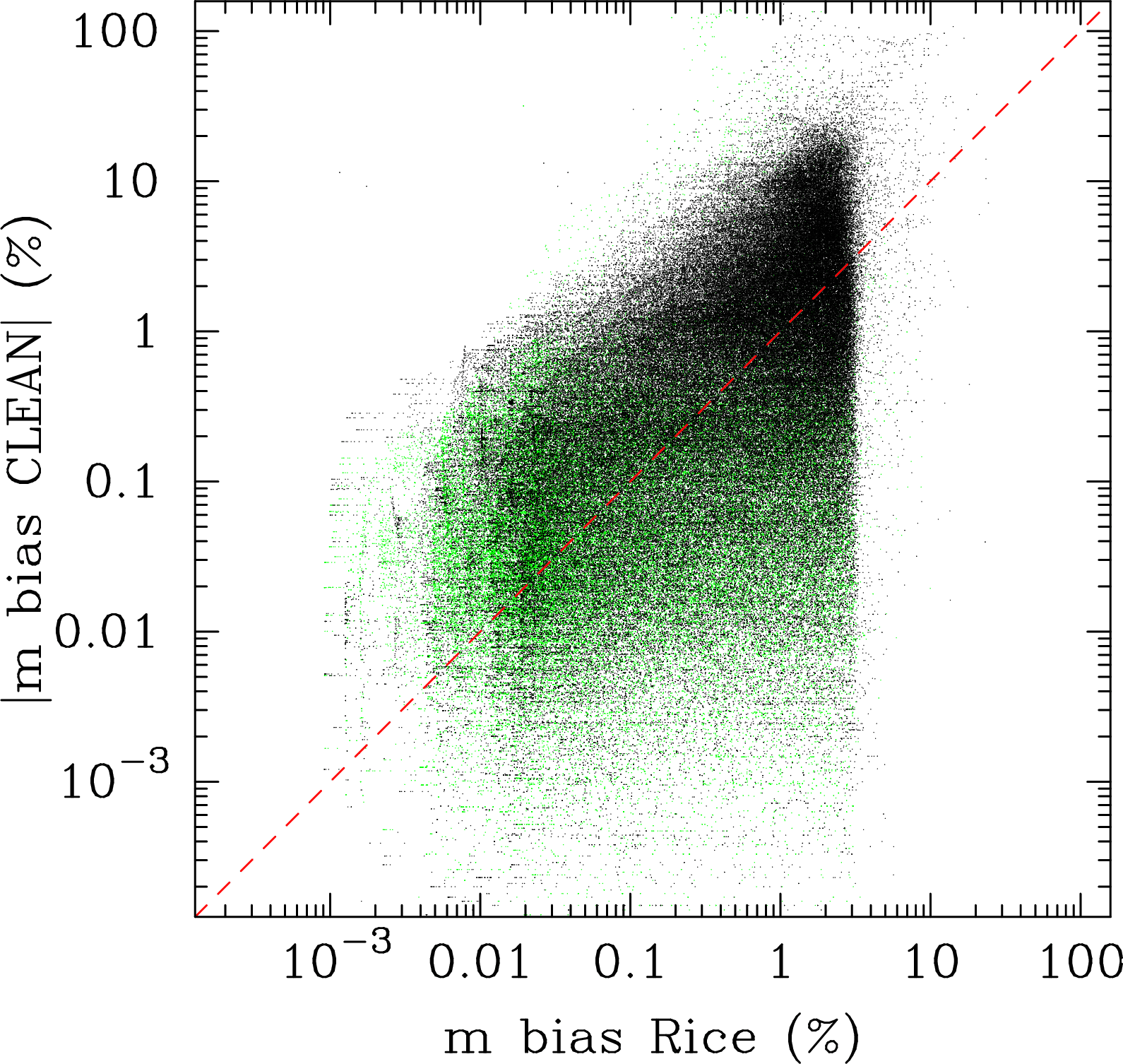}
    \caption{Left: Ratio of the absolute value of $m$-CLEAN-bias to the random error in fractional polarization as a function of $P$. The blue dots represent polarization-sensitive pixels in the core area in stacked maps, while the black points show jet regions. Right: Ratio of the absolute value of $m$-CLEAN-bias to $m$-Ricean-bias vs $m$. The green dots denote the pixels with negative values of the CLEANing bias.
    \label{f:m_clean_bias}
    }
\end{figure*}

The comparison of the absolute value of the bias in $m$ and the corresponding $\sigma_m$ both derived from simulations are given in \autoref{f:m_clean_bias}, left. It shows that the ratio $|m_{\rm \,bias\,CLEAN}|/\sigma_m$ is mainly distributed in a range from about 0.1 to several units, with median values of about 0.4 and 1.1 for the core and jet area, respectively. The polarized cores typically have the CLEANing $m$-bias smaller than the random error, while the low-SNR outer jet regions are characterised by $|m_{\rm \,bias\,CLEAN}|/\sigma_m>1$. Being compared to the Ricean bias, the CLEANing bias in fractional polarization progressively dominates towards the larger values in both of them (\autoref{f:m_clean_bias}, right), i.e. in the outer jet regions, and especially closer to the jet edges, where the biases reach 3.2~per~cent and up to 15~per~cent, respectively, in pixels with the applied polarization $\mathrm{SNR}=4$ cutoff level. Thus, is it important to correct linear polarization maps not only to the Ricean but also to the CLEANing bias. In 29~per~cent of $P$-sensitive pixels, the CLEANing $m$-bias is negative and relatively small, usually appearing in the core or inner jet regions (green dots).

We also checked whether the bias could be compensated using CLEAN models. For this, we employed simulations with a relativistic jet model with various polarization morphology. It turns out that if the distribution of the complex polarization is rather smooth (e.g. if EVPA changes smoothly) then the bias correction using our estimates works fine.

\bsp
\label{lastpage}
\end{document}